\documentclass[preprint,12pt,authoryear]{elsarticle}

\pdfoutput=1

\usepackage{amsmath}
\usepackage{bbm}

\usepackage{amssymb}
\usepackage[colorlinks = true,
            linkcolor = blue,
            urlcolor  = blue,
            citecolor = blue,
            anchorcolor = blue]{hyperref}
\usepackage{bm}
\usepackage{color}
\usepackage{hyperref}
\usepackage{caption}
\usepackage{subcaption}
\usepackage[labelformat=simple]{subcaption}
\usepackage{tabularx}
\usepackage{natbib}
\usepackage[titletoc,toc,title]{appendix}
\journal{IJSS}

\begin{document}

\begin{frontmatter}


  \title{The generalized plane piezoelectric problem: Theoretical formulation and application to heterostructure nanowires}
  \author{H. T. Mengistu and Alberto Garc\'{\i}a-Crist\'{o}bal}
  \address{Instituto de Ciencia de Materiales (ICMUV), Universidad de Valencia, E-46980 Paterna (Valencia), Spain}

\begin{abstract}

We present a systematic methodology for the reformulation of a broad class of three-dimensional (3D)
piezoelectric problems into a two-dimensional (2D) mathematical form.
The sole underlying hypothesis is
that the system geometry and material properties as well as the applied loads (forces and charges) and boundary
conditions are translationally invariant along some direction.
This requisite holds exactly in idealized indefinite systems and to
a high degree of approximation, in the sense of Saint-Venant's principle, in finite but slender systems.
This class of problems is commonly denoted here as the \emph{generalized plane piezoelectric} (\emph{GPP})
problem. For non-piezoelectric systems, the problem becomes purely elastic and
is then called the \emph{generalized plane strain} (\emph{GPS}) problem.
The first advantage of the generalized plane problems is that they are
more manageable from both analytical and computational points of view.
Moreover,
they are flexible enough to accommodate any geometric cross section,
crystal class symmetry, axis orientation and a wide range of
boundary conditions.
As an illustration we present numerical simulation of indefinite
lattice-mismatched core-shell nanowires made of diamond Ge/Si and zincblende piezoelectric
InN/GaN materials.
The remarkable agreement with exact 3D simulations of finite versions of
those systems reveal the GPP approach as a reliable procedure to study accurately
and with moderate computing resources the strain and electric field distribution in
elongated piezoelectric systems.

\end{abstract}

\begin{keyword}
Piezoelectricity, Two-dimensional approximation, Coherent inclusion, Core/shell nanowires

\end{keyword}

\end{frontmatter}

\section{Introduction}\label{INTROPAPER}

In order to analyze the piezoelectric behavior of materials, it is necessary to solve a set of coupled mechanical and electrical equations. Among the many situations where one is faced with this problem there stands out the research on semiconductors nanostructures, whose piezoelectric properties can be used to advantage in multiple applications \citep{Wang2012}. A subfield of special interest is that of pseudomorphic semiconductor heterostructures where the lattice mismatch induces elastic and piezoelectric fields in the system even in the absence of external forces or charges  \citep{Voon2011}. In any case, analytical solutions to fully-coupled piezoelectric problems in three-dimensional (3D) systems exist only under very restrictive assumptions on their geometry. On the other hand, the numerical solutions of discretized piezoelectric equations, while possible, are in general computationally expensive, especially when repeated calculations are required. One typical strategy to avoid these limitations is to approximate the 3D problem in question into a model amenable to a mathematically two-dimensional (2D) formulation, much easier to deal with it from both analytical and numerical points of view. The simplest example in the context of continuum elasticity is the standard plane strain approximation for systems translationally invariant along some direction (here conventionally taken as the $X_3$ axis), in which it is assumed that the axial displacement $u_3$ vanishes
and the other displacement  components $(u_1,u_2)$ depend only on in-plane coordinates $(x_1,x_2)$ \citep{Sadd2005}. This idea has been applied also to piezoelectric problems by further assuming that also the piezoelectric potential depends only on $(x_1,x_2)$ \citep{Rajapakse1997}. We shall use here the term \emph{plane piezoelectric} problem to refer to this situation. It has been employed by many authors:  The 2D problem of an isotropic piezoelectric material with an elliptic hole is discussed by \cite{Sosa1996}; exact solutions for the latter system subjected to uniform remote loads are obtained by \cite{Gao1999} and \cite{Dai2005}; \cite{Sosa1991} made a 2D analysis of a transversely isotropic piezoelectric solid containing defects; \cite{Chung1996} studied the 2D problem of an anisotropic piezoelectric material with an elliptic inclusion or hole using the Stroh formalism. However, the plane piezoelectric approach has limitations: there are many problems involving
specific crystal structures, orientations and loading conditions, where the medium develops out-of-plane axial $(\varepsilon_{33})$ and/or shear $(\varepsilon_{13}$ and $\varepsilon_{23})$ strain components and/or axial electric field component $(E_3)$,
that cannot be captured by the above approximation.

In this paper, we rigorously define a class of 3D fully-coupled piezoelectric problems that can be  reduced in a systematic manner to a 2D mathematical
formulation. Essentially, they correspond to systems where the geometry, the material properties, the applied loads (forces and charges), and the boundary conditions are invariant along $X_3$ axis.
This translates into the strain and electric field components being dependent only on the in-plane coordinates $(x_1,x_2)$. Under this sole hypothesis, the original 3D problem can be reformulated into a 2D mathematical framework, so that one only needs to study the cross section of the system, with the ensuing significant reduction in the computing resources needed. This approach requires no additional assumption which implies a remarkable flexibility to treat different situations, going far beyond the plain strain approximation. This set of problems will be commonly denoted in this paper as the \emph{generalized plane piezoelectric} (\emph{GPP}) problem.
The efforts to go beyond the plane strain approximation go back
to the pioneering works of Lekhnitskii in the 1930s, which were later compiled in his monograph \citep{Lekhnitskii1963}.
More recently, \cite{barber2007} have extended the Stroh formalism
to provide a class of three-dimensional solutions that have polynomial dependence on $x_3$.
The quadratic in $x_3$ displacement solutions that we study in this paper should emerge
as a particular case of the above general situation,
but it is not worked out in detail by \cite{barber2007}.
Other studies where $\varepsilon_{ij}$ is allowed to exhibit some polynomial
dependence along $x_3$-direction can be found in \citep{Lekhnitskii1963,iesan2008classical}.
The analytical solution technique introduced by \cite{barber2007} consists of a formal recursive procedure
using successive partial integrations in the $x_3$-direction. The solutions involve a sequence of matrix algebra
operations that, though routine, lead to rather complicated expressions. This method is not explicitly developed for the case of
piezoelectricity. \cite{chen1997exact} have tackled the same piezoelectric problem as in this paper
and characterize it by the same pattern of field dependencies that we propose.
However, they later proceed by a stress (Lekhnitskii) formulation,
as opposed to the displacement formulation presented here, and focus on various particular issues.
Besides its academic interest as a model for ideal infinite translationally invariant systems, it is here claimed and numerically illustrated that the GPP approach also gives useful approximate solutions, in the spirit of Saint-Venant's
principle \citep{iecsan1987saint,horgan1989recen}, for the central part of finite 3D systems with high aspect-ratio, where the deformation and electric field are essentially uniform along the axis. One of the areas where the GPP problem can be applied is in the study of  heterostructure (core-shell) nanowires (NWs), where the lattice mismatch between the constituent materials induces not only a strain distribution but also a piezoelectric field. This system has been studied recently by \cite{Boxberg2012} using direct 3D calculations.

The outline of the article is as follows. The theoretical formulation of a general 3D piezoelectric problem and the necessary prescriptions to study a lattice-mismatched piezoelectric inclusion problem are described in Sec.~\ref{3D_PIEZO}.
The hypothesis and the systematic development of the formalism underlying the 2D GPP approach are presented in Sec.~\ref{GPP-problem}. The GPP methodology is illustrated in Sec.~\ref{result} by showing  numerical results and discussions of the strain and the electric fields in   lattice-mismatched core-shell nanowires. Section \ref{Summary and conclusions} concludes the paper by summarizing the relevant accomplishments.

\section{The three-dimensional piezoelectric problem}\label{3D_PIEZO}

\subsection{General formulation of the piezoelectric problem}\label{GEN}

To fix the theoretical framework and the notation adopted, we summarize first the general formulation of a piezoelectric continuum  problem. To specify the necessary tensors we shall use index notation throughout the paper. The Latin indices $(i,j,k,l,m,n=1,2,3)$ in the tensorial objects will label the components with respect to a Cartesian reference frame $O\,X_1 X_2 X_3$, with associated coordinates $(x_1,x_2,x_3)$. Einstein summation convention applies unless the contrary is explicitly stated.

\begin{figure}[ht]
\centering
\includegraphics[width=1\textwidth]{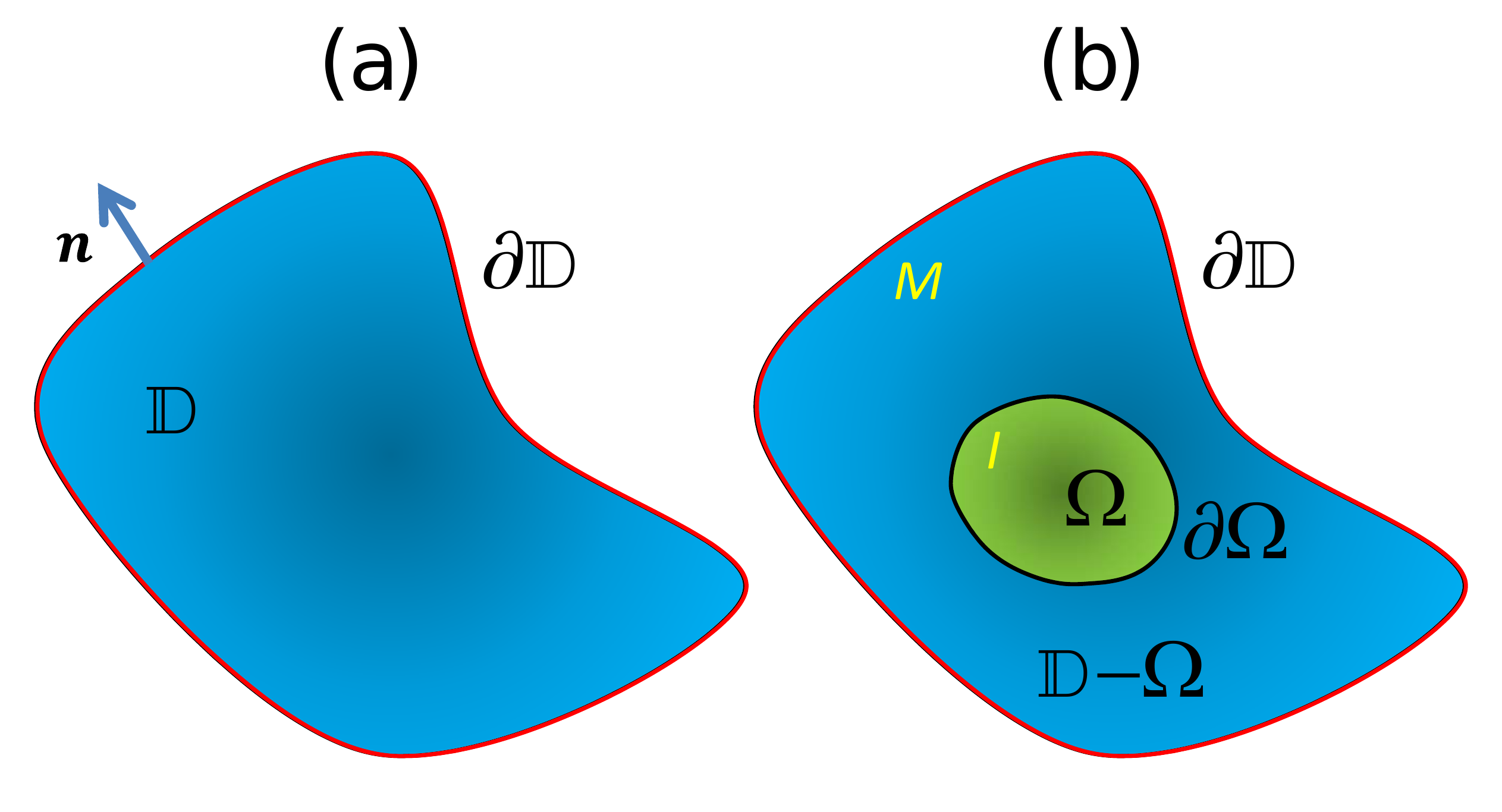}
\caption{(a) 3D piezoelectric body. (b)  Piezoelectric bimaterial system with domains $\mathbb{D}-\Omega$ (matrix $M$) and $\Omega$ (inclusion $I$), separated by the interface boundary $\partial \Omega$.  }
\label{piezoelectric_body}
\end{figure}

Let us consider a piezoelectric solid (\,see Fig. \ref{piezoelectric_body}(a)\,) that occupies a 3D domain $\mathbb{D}$ delimited by the boundary $\partial \mathbb{D}$, containing the free volume charge density $\rho$ and subjected to a body force per unit volume $f_i$. The goal is to find the distribution of
the elastic displacement vector $u_i$ and the piezoelectric potential $\phi$ over the solid. When assuming the small-deformation and electrostatic approximations, the above quantities can be related to the strain tensor $\varepsilon_{ij}$ and the piezoelectric field $E_{m}$ by the expressions:
\begin{subequations}\label{kinematical}
\begin{equation} \label{kinematical:strain}
\varepsilon_{ij}= \frac{1}{2}\left( \frac{\partial
u_i}{\partial x_j}+\frac{\partial u_j}{\partial x_i}
\right) \: ,
\end{equation}
\begin{equation}\label{kinematical:field}
E_{m}=-\frac{\partial \phi}{\partial x_{m}} \: .
\end{equation}
\end{subequations}

We restrict ourselves to the linear piezoelectric regime,
which allows to express the stress tensor $\sigma_{ij}$ and the
piezoelectric displacement vector $D_{m}$ in terms of
$\varepsilon_{ij}$ and $E_{m}$ by means of
the standard linear constitutive relations \citep{Ting1996,Hwu2010}:
\begin{subequations}\label{Constitutive}
\begin{equation}\label{EQ:Constitutive coupled_1}
\sigma_{ij}=C_{ijkl}\varepsilon_{kl}-e_{nij}E_{n} \: ,
\end{equation}
\begin{equation}\label{EQ:Constitutive coupled_2}
D_{m}=e_{mkl}\varepsilon_{kl}+\epsilon_{mn}E_{n} \: ,
\end{equation}
\end{subequations}
where $C_{ijkl}$ is the elastic stiffness tensor,
$e_{nij}$ is the piezoelectric tensor, and $\epsilon_{mn}$ is the dielectric tensor.

The equilibrium configuration
is determined by the following set of coupled differential
equations:
\begin{subequations}\label{EQ:Equlibrium}
\begin{equation}
\label{EQ:Equ1}
\frac{\partial \sigma_{ij} }{\partial x_i}=-f_j \: ,
\end{equation}
\begin{equation}
\label{EQ:Equ2}
\frac{\partial  {D}_m }{\partial x_m}=\rho \: .
\end{equation}
\end{subequations}
The first equation is the mechanical equilibrium equation,
and the second is the electrostatic Poisson equation.

When dealing with specific problems, the tensors appearing
in the general formulation of the piezoelectric problem are
often transformed into a matrix form by means of the Voigt notation \citep{Nye1985}.
In Appendix \ref{APP:Appendix A} we give the elastic, piezoelectric and dielectric matrices, $C_{IK}$, $e_{nI}$ and $\epsilon_{mn}$ ($I,K=1,\dots,6$, $m,n=1,2,3$), for the crystalline  materials belonging to the cubic system (crystal classes $T$ and $T_d$).

Equations \eqref{kinematical}-\eqref{EQ:Equlibrium}, together with appropriate boundary
conditions specified at the surface $\partial \mathbb{D}$ (with  outward normal vector $n_{i}$),
constitute the complete
mathematical description of the 3D fully-coupled piezoelectric problem.
The most general setting of boundary conditions would allow to specify either
the applied traction force
$\bar t_{j}$ or a prescribed displacement $\bar u_{j}$, and
the impressed surface charge density $\bar \varrho$ or a
fixed potential $\bar{\phi}$, in the following way:
\begin{subequations}\label{EQ:BCS}
\begin{equation}\label{EQ:BCS1}
\hphantom{-}n_{i}\sigma_{ij}= \bar t_{j}  \quad  \mathrm{on} \quad  \partial \mathbb{D} _{t} \qquad \mathrm{and} \qquad  u_{j}=\bar u_{j}
\quad  \mathrm{on} \quad \partial \mathbb{D} _{u} \: ,
\end{equation}
\begin{equation}\label{EQ:BCS11}
-n_{m}D_{m}=\bar \varrho  \quad  \mathrm{on} \quad \partial \mathbb{D}_{\varrho} \quad \mathrm{and}
   \qquad \
   \phi= \bar{\phi} \quad  \mathrm{on} \quad  \partial \mathbb{D}_{\phi} \: ,
\end{equation}
\end{subequations}
where $(\partial \mathbb{D} _{t},\partial \mathbb{D} _{u})$ and $(\partial \mathbb{D} _{\varrho},\partial \mathbb{D} _{\phi})$ represent two, in general different, partitions of the boundary $\partial \mathbb{D}$. The physical problems are usually modeled by a simpler situation,
the simplest one being the
uncharged free boundary (that would correspond to $\partial \mathbb{D} _{t}=\partial \mathbb{D}_{\varrho}=\partial \mathbb{D}$ with $\bar
t_{j}=0$ and $\bar \varrho=0$).

The so-called semi-coupled approach to the piezoelectric problem consists of neglecting
the piezoelectric contribution to the stress, by imposing $e_{nij}\to 0$ in Eq.
\eqref{EQ:Constitutive coupled_1}, and solving the resulting purely mechanical problem
given by Eqs. \eqref{EQ:Equ1} and
\eqref{EQ:BCS1}.
In a second decoupled step, the obtained strain $\varepsilon_{ij}$ is inserted into Eq.\eqref{EQ:Constitutive coupled_2}
and the electrostatic Poisson problem given by Eqs. \eqref{EQ:Equ2} and
\eqref{EQ:BCS11}
is solved to give the piezoelectric field $E_n$ and the potential $\phi$.
Of course, in the case of a non-piezoelectric material the piezoelectric constants vanish
exactly ($e_{nij}=0$) in every expression, and one has to solve separately the
uncoupled
mechanical and electrostatic problems.

We remind here that the piezoelectric problem just stated can be deduced from
the principle of virtual work.
This principle asserts that
the sum of the work from internal stresses $\sigma_{ij}$ and electric displacements $D_m$
and the external work $\delta W_\mathrm{ext}$, during an admissible virtual displacement $\delta u_i$ and
potential variation $\delta\phi$ around the physical equilibrium solutions $u_i$ and $\phi$, is zero
\citep{tiersten1969book}:
\begin{equation}\label{Var1}
  -\int_\mathbb{D} d^{\, 3}\bm{r}\,\left\{ \sigma _{ij} \,\frac{\partial \delta u_j}{\partial x_i}
 +D_m\,\frac{\partial \delta \phi}{\partial x_m}\right\}+\delta W_\mathrm{ext}=0
  \quad .
\end{equation}
In our case the external work is to be written as:
\begin{equation}\label{Var2}
\delta W_\mathrm{ext} =\int_\mathbb{D} d^{\, 3}\bm{r}\,\left( f_j\,\delta u_j-\rho\,\delta\phi\right)+
\int_{\partial \mathbb{D}_t} dS \, \bar{t}_j\,\delta u_j-\int_{\partial \mathbb{D}_\varrho} dS \,\bar{\varrho}\,\delta\phi \quad .
\end{equation}
The integral formulation in Eqs.~\eqref{Var1} and \eqref{Var2}
can be used as a starting point to generate practical and flexible numerical approximation methods,
such as the popular finite element method (FEM).

\subsection{The coherent piezoelectric inclusion problem}\label{INC}

One problem of particular interest
is that of finding the elastic and electric fields induced in
a coherent (or pseudomorphic) lattice-mismatched bimaterial system.
Such a heterostructure consists of two
domains, $\mathbb{D}-\Omega$ and $\Omega$, occupied by
two materials that have the same crystalline structure but differ in their lattice parameters
(\,see Fig.~\ref{piezoelectric_body}(b)\,). Quite conventionally these domains are respectively called the \emph{matrix} (\,associated quantities will be hereafter labeled with $(M)$\,) and the \emph{inclusion} (\,label $(I)$\,).
Their lattice parameters are denoted by $a_i^{(M)}$ and $a_i^{(I)}$, $i=1,2,3$.
The contact interface between both materials is assumed to be coherent, i.e., dislocation-free, despite the existing lattice mismatch \citep{povolotskyi2006elasticity}. This requirement is the cause for the appearance of a certain strain and field distribution over the system, that
we want to calculate.

For later reference it is convenient to introduce here the
so-called \emph{misfit} (or  \emph{mismatch}) \emph{strain} derived from the nominal lattice mismatch
between the matrix  and  inclusion materials:
\begin{equation}\label{mis-fit}
\varepsilon^{(\mathrm{misfit})}_{ij}= \varepsilon^{(\mathrm{misfit})}_{i}\,\delta_{ij}
\leftrightarrow
\begin{pmatrix}
\varepsilon^{(\mathrm{misfit})}_1 & 0 & 0   \\
0 & \varepsilon^{(\mathrm{misfit})}_2 &0 \\
0 &
0 &
\varepsilon^{(\mathrm{misfit})}_3
\end{pmatrix}
\:  ,
\end{equation}
with
\begin{equation}
\varepsilon^{(\mathrm{misfit})}_i
=
\frac{a_i^{(M)}-a_i^{(I)}}{a_i^{(I)}}\quad \quad (i=1,2,3) \quad .
\end{equation}
Note that in Eq.~\eqref{mis-fit}, and in Eq.~\eqref{mis-fit-ref} below,
the repeated index $i$ is not summed.

The elastic constants of the heterostructure can be written as:
\begin{equation}\label{C_chi}
C_{ijkl}(\bm{r})=C_{ijkl}^{(M)}\chi^{(M)}(\bm{r}) + C_{ijkl}^{(I)}\chi^{(I)}(\bm{r})\: ,
\end{equation}
where
$\chi^{(I)}$ is the characteristic function of the inclusion defined as:
\begin{equation}\label{Chi}
\chi^{(I)}(\bm{r})=
\left\{
\begin{array}{ccl}
1 & \quad & {\rm if \ } \bm{r}\in\Omega \\
0 & \quad & {\rm if \ } \bm{r}\in \mathbb{D}-\Omega
\end{array}
\right. \: ,
\end{equation}
and $\chi^{(M)}=1-\chi^{(I)}$ is the characteristic function of the matrix.
Similar expressions to Eq.~\eqref{C_chi} can be written for the piezoelectric constants
$e_{nij}(\bm{r})$ and dielectric constants $\epsilon_{mn}(\bm{r})$
of the heterostructure.

A generalization of the classical Eshelby inclusion method,
well-known in the micromechanics literature \citep{Eshelby1961,Mura1987},
provides a systematic procedure to obtain the strain and electric
field in the above described system.
It essentially amounts to a
\textit{gedanken} procedure in which the two material domains are first independently constrained to a
common crystal lattice, characterized by reference lattice parameters $a_i^{(\mathrm{ref})}$,
by applying appropriate
stresses and charges.
For later use it is convenient to introduce here the notation:
\begin{equation}\label{mis-fit-ref}
\varepsilon^{(0)}_{ij}(\bm{r})= \frac{a_i^{(\mathrm{ref})}-a_i^{(M)}}{a_i^{(M)}}\,\delta_{ij}\,\chi^{(M)}(\bm{r})+
\frac{a_i^{(\mathrm{ref})}-a_i^{(I)}}{a_i^{(I)}}\,\delta_{ij}\,\chi^{(I)}(\bm{r})
\: .
\end{equation}
Note that if we take $a_i^{(\mathrm{ref})}=a_i^{(M)}$ then:
\begin{equation}\label{mis-fit-ref_1}
\varepsilon^{(0)}_{ij}(\bm{r})=
\varepsilon^{(\mathrm{misfit})}_{ij}\,\chi^{(I)}(\bm{r})
\: ,
\end{equation}
which is the usual choice when treating inclusions in an infinite matrix.

Following with the \textit{gedanken} procedure, the now lattice-matched
material domains are coherently joined, and left to relax to the final equilibrium configuration
under applied stresses and charges opposite to the ones in the previous step,
thereby removing any external action on the system. We summarize here the final results of this procedure:

\begin{itemize}

\item First, it has to be noticed that the Eshelby procedure gives
the total strain with respect to the undeformed state of the local lattice as the sum of two terms:
\begin{equation}\label{Total_strain}
\varepsilon^{(T)}_{ij}(\bm{r})= \varepsilon^{(0)}_{ij}(\bm{r})+
\varepsilon_{ij}(\bm{r})\: .
\end{equation}
In the context of the coherent inclusion problem, the unknown $\varepsilon_{ij}(\bm{r})$
describes the strain state attained after relaxation from the reference lattice configuration,
and therefore is here  called the \emph{relaxation strain}.
Associated to $\varepsilon_{ij}(\bm{r})$  we have a displacement
field $u_i$ as given by Eq.~\eqref{kinematical:strain} and
constitutive relations as given by Eq.~\eqref{Constitutive}.

\item  The final equilibrium configuration for the
relaxation strain $\varepsilon_{ij}(\bm{r})$ and the electric field
$E_m(\bm{r})$ can be obtained by solving the following set of
coupled partial differential equations:
\begin{subequations}\label{EQ_INC}
\begin{equation}
\label{EQ_INC:Equ1}
\frac{\partial \sigma_{ij} }{\partial x_i}=  -f_j^{(0)} \: ,
\end{equation}
\begin{equation}
\label{EQ_INC:Equ2}
 \frac{\partial  {D}_m }{\partial x_m}= \rho^{(0)} \: ,
\end{equation}
\end{subequations}
where the lattice mismatch induced  force $f_i^{(0)}$ and charge $\rho^{(0)}$ are
given by:
\begin{subequations}\label{INCLUSION}
\begin{alignat}{2}
 f_j^{(0)}=\frac{\partial \sigma^{(0)}_{ij}  }{\partial x_i}
 \quad & \mathrm{with} \quad \sigma^{(0)}_{ij}(\bm{r})= C_{ijkl}(\bm{r})\,\varepsilon^{(0)}_{kl}(\bm{r})
 \: , \label{EQ_INCLUSION:Equ1} \\[1ex]
 \rho^{(0)}=-\frac{\partial P^{(0)}_m }{\partial x_m}
 \quad & \mathrm{with} \quad
  P^{(0)}_m(\bm{r}) = e_{mkl}(\bm{r})\,\varepsilon^{(0)}_{kl}(\bm{r})\
 \: . \label{EQ_INCLUSION:Equ2}
\end{alignat}
\end{subequations}
Note that, due to the presence of the step-like characteristic functions
inside the derivatives in Eq. \eqref{INCLUSION},
$f_i^{(0)}$ and $\rho^{(0)}$ represent actually surface force and charge
applied on the interface $\partial\Omega$ separating the inclusion and matrix.

\item If the system is further loaded with force $f_i$ and/or charge
$\rho$, and/or subjected to arbitrary boundary conditions on the surface $\partial \mathbb{D}$, these
effects can be easily
added, by means of the superposition principle, to the
inclusion problem represented by Eq. \eqref{EQ_INC}.

\end{itemize}

In conclusion, we have shown that the particular problem of a coherent
piezoelectric inclusion can be mapped to a standard piezoelectric problem as
described in Sec. \ref{GEN} by a proper introduction of equivalent forces and charges.
As long as $\varepsilon^{(0)}_{ij}$ is small, the results of the generalized Eshelby
procedure will be rather insensitive to the specific choice of $a_i^{(\mathrm{ref})}$.

\normalsize
\section{The generalized plane piezoelectric (GPP) problem}\label{GPP-problem}

\subsection{Hypothesis underlying the GPP problem}\label{GPP-problem_1}

\begin{figure}[ht]
\centering
\includegraphics[width=1\textwidth]{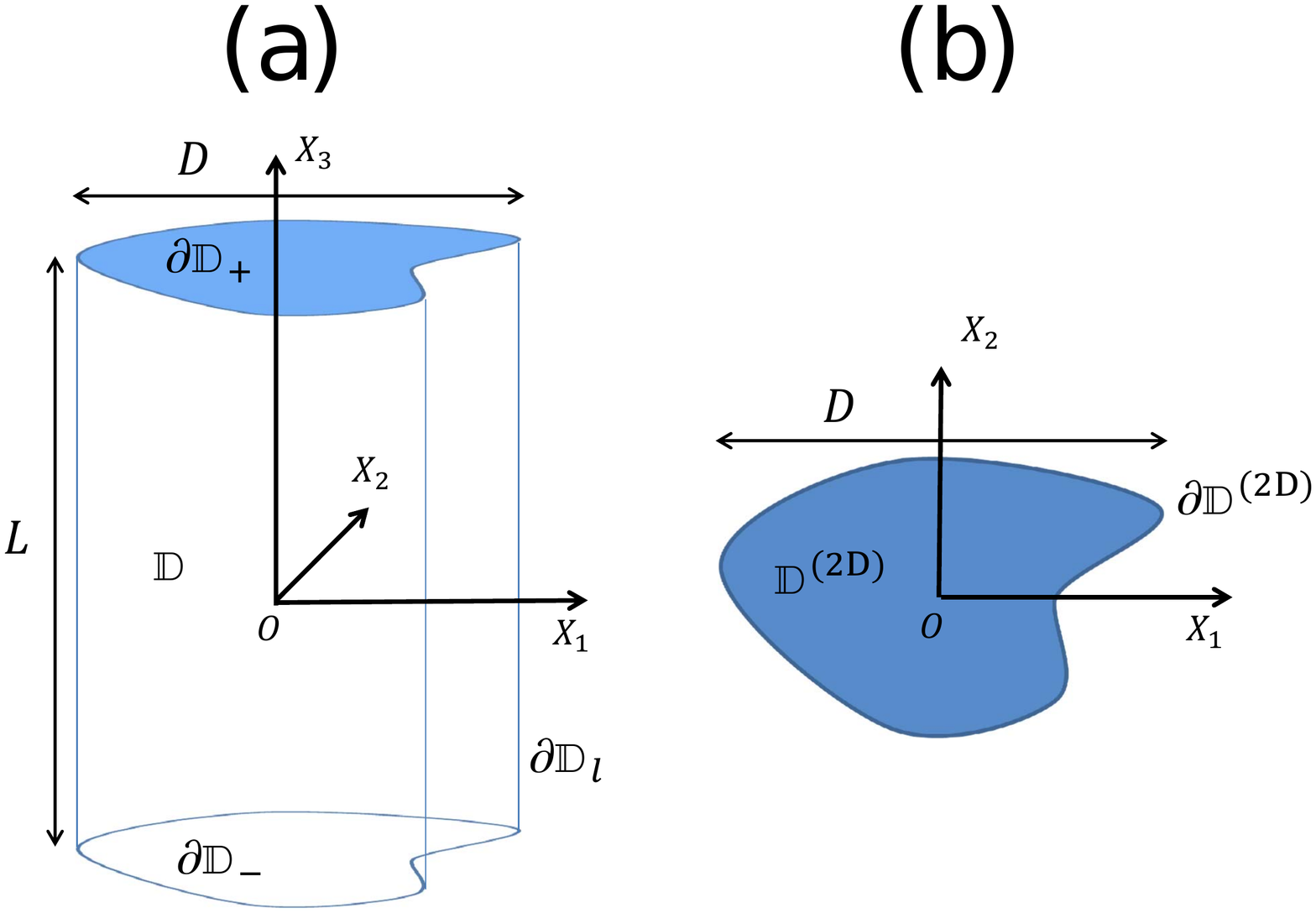}
\caption{Sketch of the geometry of the generalized plane piezoelectric problem:
(a) 3D geometry and (b) 2D cross section.}
\label{2D-approximation}
\end{figure}

As explained in Sec. \ref{INTROPAPER}, the plane piezoelectric problem
is not flexible enough to accommodate many situations.
To overcome such limitations we propose in this work a more general 2D approach,
which is called here the \emph{generalized plane piezoelectric (GPP)} problem.
More specifically, the class of problems adapted to the GPP approach to be developed below correspond to the situation sketched in Fig.~\ref{2D-approximation}(a): The geometry of the general piezoelectric body of Fig.~\ref{piezoelectric_body} is further restricted here by assuming a right  cylindrical shape, oriented along the $X_3$ axis (hereafter referred as the longitudinal axis), and with constant cross section.
The axial length of the system is in principle finite with magnitude $L$.
In this geometry, the boundary delimiting $\mathbb{D}$ can be naturally decomposed as $\partial \mathbb{D}=\partial \mathbb{D}_l\bigcup
\partial \mathbb{D}_+\bigcup
\partial \mathbb{D}_-$, where
$\partial \mathbb{D}_l$ is the lateral surface of the cylinder,
and $\partial \mathbb{D}_{\pm}$ are the two extreme sections of the
body, at $x_3=\pm \frac{L}{2}$.
The transversal section of $\mathbb{D}$ determines a 2D
domain $ \mathbb{D}^{(\mathrm{2D})}$, its boundary $\partial\mathbb{D}^{(\mathrm{2D})}$
being determined by the transversal section of $\partial \mathbb{D}_l$ (\,see Fig. ~\ref{2D-approximation}(b)\,).
In addition, we assume that the material constants are independent
of the axial coordinate $x_3$.
The system may be subjected to body or boundary loads and/or displacement restrictions, as well
as impressed charges and/or applied potentials, as explained in Sec.~\ref{GEN}, but we assume
that the quantities representing these actions
are also independent of the $x_3$ coordinate.

Even after all these assumptions, due to the end effects, the problem is still 3D, i.e.,
$\varepsilon_{ij}$ and $E_m$ depend on $(x_1,x_2,x_3)$.
If we are interested in the exact solution (particularly, the behavior near the ends of the system),
there is no option but to solve the genuine 3D problem.
However, in many cases the system has a high aspect ratio, i.e., $L\gg D$
(say $L/D\gtrsim 2-3$), where $D$ is the largest dimension of the
cross section $ \mathbb{D}^{(\mathrm{2D})}$ (\,see Fig. ~\ref{2D-approximation}(b)\,).
For such a system, Saint-Venant's principle of linear elasticity
suggests that,
far from the end sections $\partial \mathbb{D}_{\pm}$,
\emph{it is expected} that all the cross sections along the longitudinal axis can be considered to be at
identical conditions \citep{iecsan1987saint,horgan1989recen}.
Hence, the strain and electric field  distribution
at the central part of the body can be described as invariant along the longitudinal $X_3$ direction
and dependent at most on the in-plane coordinates $(x_1,x_2)$ \citep{Lekhnitskii1963,Hwu2010}:
\begin{subequations}\label{GPP}
\begin{equation}
\label{GPP_1}
\varepsilon_{ij}=\varepsilon_{ij}(x_{1},x_{2}) \, ,
\end{equation}
\begin{equation}
\label{GPP_2}
E_{m}=E_{m}(x_{1},x_{2}) \, .
\end{equation}
\end{subequations}
Note that here it is not required a priori that any strain and/or electric
field component vanishes, in contrast to the assumptions of the standard
plane piezoelectric approximation. In the following, we examine
in detail
the consequences of the ansatz \eqref{GPP}.

In the first place,
by carefully integrating the
kinematical relations \eqref{kinematical} with respect to $x_3$,
under the
constraint \eqref{GPP}, the following general expressions of the displacement field
and the electric potential are obtained \citep{chen1997exact}:
\begin{subequations}\label{EQ:displacements}
\begin{alignat}{2}
u_{1}(x_{1},x_{2},x_{3})&=U_{1}(x_{1},x_{2})-\frac{1}{2}\frac{1}{R_1}{x_{3}}^2+\theta x_{2}x_{3}   \, , \label{EQ:Field_DISPLACEMENT:Equ1} \\[1ex]
u_{2}(x_{1},x_{2},x_{3})&=U_{2}(x_{1},x_{2})-\frac{1}{2}\frac{1}{R_2}{x_{3}}^2-\theta x_{1}x_{3}   \, ,
\label{EQ:Field_DISPLACEMENT:Equ2} \\[1ex]
u_{3}(x_{1},x_{2},x_{3})&=U_{3}(x_{1},x_{2})+\varepsilon_{\parallel} x_{3}+\frac{1}{R_1}x_{1}x_{3}
+\frac{1}{R_2}x_{2}x_{3} \, ,
\label{EQ:Field_DISPLACEMENT:Equ3}
\end{alignat}
\end{subequations}
\begin{equation}\label{EQ:Field_DISPLACEMENT}
\phi(x_{1},x_{2},x_{3})=\Phi(x_{1},x_{2})-E_{\parallel}x_{3} \, ,
\end{equation}
where $\varepsilon_{\parallel}$, $R_1$, $R_2$, $\theta$, and $E_{\parallel}$ are constants, and
$U_i(x_1,x_2)$ ($i=1,2,3$) and $\Phi(x_1,x_2)$ are
mathematically 2D fields.
The expressions \eqref{EQ:displacements} and \eqref{EQ:Field_DISPLACEMENT}
represent a generalization of those introduced by Lekhnitskii for linear elastic materials
\citep{Lekhnitskii1963}.
By introducing Eqs.~\eqref{EQ:displacements} and \eqref{EQ:Field_DISPLACEMENT} into Eq. \eqref{kinematical},
we can also obtain the
general form of the strain tensor:
\begin{subequations}\label{EQ:strains}
\begin{equation}\label{EQ:strains_gen}
\varepsilon_{ij}(x_{1},x_{2})=\varepsilon^{(U)}_{ij}(x_{1},x_{2})
+\varepsilon^{(\bullet)}_{ij}(x_{1},x_{2})  \, ,
\end{equation}
\begin{alignat}{2}
\label{EQ:strain_inplane}
\varepsilon^{(U)}_{ij}(x_{1},x_{2}) & \leftrightarrow
\begin{pmatrix}
\frac{\partial U_{1}}{\partial x_{1}}     &
\frac{1}{2}(\frac{\partial U_{1}}{\partial x_{2}}+\frac{\partial U_{2}}{\partial x_{1}}) &
\frac{1}{2}\frac{\partial U_{3}}{\partial x_{1}}   \\
\times &
\frac{\partial U_{2}}{\partial x_{2}} &
\frac{1}{2}\frac{\partial U_{3}}{\partial x_{2}} \\
\times &
\times &
0
\end{pmatrix}
 \, ,
\\[1ex]
\label{EQ:strain_outoffplane}
\varepsilon^{(\bullet)}_{ij}(x_{1},x_{2})& \leftrightarrow
\begin{pmatrix}
0 & 0 &\frac{1}{2}\theta x_{2}   \\
0 & 0 &-\frac{1}{2}\theta x_{1} \\
\times &
\times &
\varepsilon_{\parallel}+\frac{1}{R_1}x_{1} +\frac{1}{R_2}x_{2}
\end{pmatrix}
\, ,
\end{alignat}
\end{subequations}
and the electric field:
\begin{equation}
\label{EQ:fields}
\bm{E}(x_{1},x_{2})=
\begin{pmatrix}
-\frac{ \partial \Phi }{ \partial x_{1}  }     \\
-\frac{  \partial\Phi }{ \partial x_{2}  }     \\
0
\end{pmatrix}
+\begin{pmatrix}
0  \\
0  \\
E_\parallel
\end{pmatrix}
\equiv \bm{E}^{(\Phi)}(x_{1},x_{2})+E_\parallel\bm{u}_3 \, .
\end{equation}
The symbol $\times$ in Eqs. \eqref{EQ:strain_inplane} and \eqref{EQ:strain_outoffplane} means
that the corresponding matrix elements are obtained from the symmetry of the strain tensor.
Moreover, the inspection of \eqref{EQ:strains} and \eqref{EQ:fields} provides a clear
interpretation of
the different constants and fields introduced in
Eqs. \eqref{EQ:displacements} and \eqref{EQ:Field_DISPLACEMENT}.
Thus, the vector field
$(U_1,U_2,U_3)$  represents the part of the displacement which is
invariant along the $X_3$ axis. The out-of-plane displacement
$U_3$  is commonly called the warping function.
Further,  $\Phi$  is the part of the potential leading to the
in-plane projection of the field. It will be called hereafter the
in-plane piezoelectric potential.
The constants $\varepsilon_{\parallel}$, $R_1$, $R_2$, $\theta$, and $E_{\parallel}$
have the following meaning:
\begin{itemize}
\item   $\varepsilon_{\parallel}$  is the axial strain describing the relative elongation of the of the system along the $X_3$ axis.
\item   $R_1$ ($R_2$) is the curvature radius associated to the bending of the body in the $X_{1}X_{3}$ ($X_{2}X_{3}$) plane.
\item  $\theta$ is the twist per unit length associated to the torsion of the body about the $X_3$ axis.
\item  $E_{\parallel}$ is the electric field along the $X_3$ axis.
\end{itemize}

The general form of the stress tensor and electric displacement field
compatible with the ansatz \eqref{GPP}
are obtained by combining Eqs. \eqref{EQ:strains} and \eqref{EQ:fields}
with the constitutive relations \eqref{Constitutive} to obtain:
\begin{subequations}\label{EQ:stress}
\begin{equation}
\sigma_{ij}(x_{1},x_{2})=\sigma^{(U\Phi)}_{ij}(x_{1},x_{2})
+\sigma^{(\bullet)}_{ij}(x_{1},x_{2})  \, ,
\end{equation}
\begin{alignat}{2}
\sigma^{(U\Phi)}_{ij}(x_{1},x_{2})&=
C_{ijkl}\,\varepsilon^{(U)}_{kl}(x_{1},x_{2})
-e_{nij}\,E^{(\Phi)}_{n}(x_{1},x_{2}) \, ,
\\[1ex]
\sigma^{(\bullet)}_{ij}(x_{1},x_{2})&=
C_{ijkl}\,\varepsilon^{(\bullet)}_{kl}(x_{1},x_{2})
-e_{nij}\,E_\parallel\delta_{n3} \, .
\end{alignat}
\end{subequations}

\begin{subequations}\label{EQ:Dfield}
\begin{equation}
D_{m}(x_{1},x_{2})=D^{(U\Phi)}_{m}(x_{1},x_{2})
+D^{(\bullet)}_{m}(x_{1},x_{2})  \, ,
\end{equation}
\begin{alignat}{2}
D^{(U\Phi)}_{m}(x_{1},x_{2})&=
e_{mkl}\,\varepsilon^{(U)}_{kl}(x_{1},x_{2})
+\epsilon_{mn}\,E^{(\Phi)}_{n}(x_{1},x_{2}) \, ,
\\[1ex]
D^{(\bullet)}_{m}(x_{1},x_{2})&=
e_{mkl}\,\varepsilon^{(\bullet)}_{kl}(x_{1},x_{2})
+\epsilon_{mn}\,E_\parallel\delta_{n3} \, .
\end{alignat}
\end{subequations}
In Eqs.~\eqref{EQ:stress} and \eqref{EQ:Dfield}, the labels ${(U\Phi)}$ and ${(\bullet)}$
refer to those parts of $\sigma_{ij}$ and $D_m$ that depend on the fields $(U_1,U_2,U_3,\Phi)$
and the constants $(\varepsilon_{\parallel}, R_1, R_2,$ $\theta, E_{\parallel})$, respectively.

To sum up, the condition \eqref{GPP} has been shown to determine the most general form of the various fields as expressed in detail by Eqs. \eqref{EQ:displacements}-\eqref{EQ:Dfield}. The piezoelectric problem that complies with that condition and the consequent fields pattern is here called a \emph{generalized plane piezoelectric (GPP)} problem.
If the materials involved are not piezoelectric (i.e., $e_{nij}=0$) we would encounter
uncoupled \emph{generalized plane strain (GPS)} and \emph{generalized plane electrostatic} problems.
 Note that in a GPP problem
the strain and electric field are independent of $x_3$, but the mechanical displacement
and electric potential can depend on $x_3$ as well as on $(x_1,x_2)$.
If one further requires that $\frac{\partial u_i}{\partial x_3}=0=\frac{\partial \phi}{\partial x_3}$,
then the standard plane piezoelectric problem is recovered.
It is worth mentioning that other authors have used the \emph{generalized plane} qualification
for problems with a more restricted scope than our defining condition \eqref{GPP} (\,see \cite{cheng1995generalized,kotousov2003generalized,li2005variational}\,).
Those problems can always be treated as particular cases of the general situation described in this paper.

\subsection{Equilibrium equations for the GPP problem}

So far, we have specified the structure of the fields for the GPP problem.
It is necessary now to establish the corresponding form of the equilibrium equations.

First, we introduce the GPP form of the stress \eqref{EQ:stress}
and electric displacement \eqref{EQ:Dfield} into the general 3D equilibrium equations \eqref{EQ:Equlibrium}
to obtain:
\begin{subequations}\label{pre-Navier-Poisson}
\begin{equation}
\frac{\partial \sigma^{(U\Phi)}_{\alpha j} }{\partial x_\alpha}+\frac{\partial \sigma^{(\bullet)}_{\alpha j} }{\partial x_\alpha}
=-f_j(x_1,x_2)  \, ,
\end{equation}
\begin{equation}
 \frac{\partial  {D}^{(U\Phi)}_\alpha }{\partial x_\alpha}+\frac{\partial  {D}^{(\bullet)}_\alpha }{\partial x_\alpha}=\rho(x_1,x_2)   \, .
\end{equation}
\end{subequations}
Hereafter, the Latin indices continue to run over all spatial directions, i.e., $i,j,k,l,m,n= 1,2,3$,
whereas Greek indices will run only over in-plane directions, i.e., $\alpha,\beta=1,2$.
Note that $\sigma_{33}$ and $D_3$ do not appear in the equilibrium equations, since they are determined
by the remaining components, as can be shown by making use of the inverse constitutive equations relating
$\varepsilon_{33}$ and $E_3$ to $\sigma_{ij}$ and $D_m$.

Finally, after inserting  Eqs. \eqref{EQ:strains} and \eqref{EQ:fields}
into Eqs. \eqref{EQ:stress} and \eqref{EQ:Dfield},
one gets expressions for the stress tensor and electric displacement vector
in terms of the fields $U_i$ and $\Phi$. These expressions can be entered into
Eq. \eqref{pre-Navier-Poisson} to produce the GPP problem equilibrium
equations, that read in a compact matrix form as:
\begin{equation}\label{EQ:Navier_poisson}
\begin{pmatrix}
\hat{L}_{11} & \hat{L}_{12} &\hat{L}_{13} & \hat{L}_{14} \\
\hat{L}_{21} & \hat{L}_{22} & \hat{L}_{23} & \hat{L}_{24}  \\
\hat{L}_{31} & \hat{L}_{32} & \hat{L}_{33} & \hat{L}_{34} \\
\hat{L}_{41}   &\hat{L}_{42}   & \hat{L}_{43}   & \hat{L}_{44} \\
\end{pmatrix}
\begin{pmatrix}
U_{1} \\
U_{2} \\
U_{3} \\
\Phi \\
\end{pmatrix}
=
\begin{pmatrix}
-f^{(\bullet)}_1 \\
-f^{(\bullet)}_2 \\
-f^{(\bullet)}_3 \\
\rho^{(\bullet)} \
\end{pmatrix} \quad ,
\end{equation}
where the matrix elements  $\hat{L}_{jk}$ are linear differential operators defined as:
\begin{equation}
 \begin{split}
 \label{EQ:PDE_operator}
\hat{ L}_{jk}=\frac{{\partial} }{\partial {x_{\alpha}} }\, C_{\alpha j \beta k}\,\frac{{\partial} }{\partial {x_{\beta}} } \quad , \\
\hat{L}_{j4}= \frac{{\partial} }{\partial {x_{\alpha}} }\,e_{\beta,\alpha j}\,\frac{{\partial} }{\partial {x_{\beta}} } \quad , \\
  \hat{L}_{4k}= \frac{{\partial} }{\partial {x_{\alpha}} }\,e_{\alpha,\beta k}\,\frac{{\partial} }{\partial {x_{\beta}} } \quad ,\\
   \hat{L}_{44}=-\frac{{\partial} }{\partial {x_{\alpha}} }\,\epsilon_{\alpha\beta}\,\frac{{\partial} }{\partial {x_{\beta}} } \quad , \\
  \end{split}
\end{equation}
the inhomogeneous terms $f^{(\bullet)}_j$
are modified body forces given by
\begin{equation}\label{EQ:BF1}
f^{(\bullet)}_{j}=f_{j}+\frac{\partial \sigma^{(\bullet)}_{\alpha  j }}{\partial x_\alpha} \quad ,
\end{equation}
and $\rho^{(\bullet)}$ is a modified charge density given by
\begin{equation}
\label{EQ:Body_force}
\rho^{(\bullet)}=\rho-\frac{\partial  {D}^{(\bullet)}_\alpha }{\partial x_\alpha} \quad .
\end{equation}

The above equations can be applied for inhomogeneous  material properties, but remember that in the context of the GPP problem
they may depend at most on the in-plane coordinates, i.e., $C_{ijkl}(x_{1},x_{2})$, $e_{nij}(x_{1},x_{2})$ and  $\epsilon_{mn}(x_{1},x_{2})$.

Since $f^{(\bullet)}_i$ and $\rho^{(\bullet)}$ depend implicitly
on $(\varepsilon_{\parallel}, R_1, R_2, \theta, E_{\parallel})$,
the non-ho\-mo\-ge\-neous system of coupled partial differential equations \eqref{EQ:Navier_poisson}
must be solved under appropriate boundary conditions, in a self-consistent manner,
for the unknown in-plane fields $U_{i}(x_1,x_2)$
and $\Phi(x_1,x_2)$ and constants $(\varepsilon_{\parallel}, R_1, R_2,$ $\theta, E_{\parallel})$.
We note that, although not carried further on here,
the solutions of the homogeneous version of system \eqref{EQ:Navier_poisson}
can be conveniently studied by using the
Stroh formalism \citep{Ting1996,chen1997exact,Hwu2010}.

For general anisotropic piezoelectric materials,
the out-of-plane displacement (warping function) $U_3$ is coupled to the in-plane
displacements $U_\alpha$ and potential
$\Phi$.
However, in the case of materials for which $C_{I4}=0=C_{I5}$ (for $I=1,2,6$)
and $e_{\beta 4}=0=e_{\beta 5}$ (for $\beta=1,2$),
one gets that $\hat{L}_{\alpha 3}=0=\hat{L}_{3\alpha}$ ($\alpha=1,2$)
and $\hat{L}_{4 3}=0=\hat{L}_{34}$, and therefore the equilibrium equations simplify as
follows:
\begin{equation}\label{EQ:Navier_poisson2}
\begin{pmatrix}
\hat{L}_{11} & \hat{L}_{12} &0 & \hat{L}_{14} \\
\hat{L}_{21} & \hat{L}_{22} & 0& \hat{L}_{24}  \\
0 & 0 & \hat{L}_{33} & 0 \\
\hat{L}_{41}   &\hat{L}_{42}   & 0  & \hat{L}_{44} \\
\end{pmatrix}
\begin{pmatrix}
U_{1} \\
U_{2} \\
U_{3} \\
\Phi \\
\end{pmatrix}
=
\begin{pmatrix}
-f^{(\bullet)}_1 \\
-f^{(\bullet)}_2 \\
-f^{(\bullet)}_3 \\
\rho^{(\bullet)} \\
\end{pmatrix}
\quad ,
\end{equation}
the differential equation related to
the warping function $U_{3}$ being decoupled from the in-plane problem associated to $(U_1,U_2,\Phi)$.
The above requirements on the material constants hold for specific situations of interest, such
as the case of diamond- and zincblende-based systems with their longitudinal axis along the
[001] direction (see Appendix \ref{APP:Appendix A})
and wurtzite-type systems along the [0001] direction.
There are other interesting cases, such as the diamond- and zincblende-based systems
with axis along the [111] direction, that do not comply with the above material symmetry requirements
(see Appendix \ref{APP:Appendix A}) and they exhibit a warping function fully-coupled into the
piezoelectric problem as illustrated by the numerical results in Sec.\ref{result}.

As commented at the end of Sec. \ref{GPP-problem_1}, in the case of general non-piezoelectric materials  with $e_{nij}=0$,
one has to deal separately with the uncoupled 2D electrostatic and elastic problems. The electrostatic problem
amounts to solve the 2D Poisson equation. In the absence of body and surface charges or
potentials, though, one is left only with a purely elastic \emph{generalized plane strain} (\emph{GPS}) problem \citep{Blazquez2006}:
\begin{equation}
 \label{EQ:Navier_GPS}
\begin{pmatrix}
\hat{L}_{11} & \hat{L}_{12} &\hat{L}_{13}  \\
\hat{L}_{21} & \hat{L}_{22} & \hat{L}_{23}   \\
\hat{L}_{31} & \hat{L}_{32} & \hat{L}_{33}  \\
\end{pmatrix}
\begin{pmatrix}
U_{1} \\
U_{2} \\
U_{3} \\
\end{pmatrix}
\equiv
\begin{pmatrix}
-f^{(\bullet)}_1 \\
-f^{(\bullet)}_2 \\
-f^{(\bullet)}_3 \\
\end{pmatrix}\, .
\end{equation}
As before, there exists, in general, the coupling between $(U_1,U_2)$ and $U_3$. Only for
materials with $C_{I4}=0=C_{I5}$ (for $I=1,2,6$), the equilibrium equations
become uncoupled and simplify as:
\begin{equation}
 \label{EQ:GPP_MATRIX4}
\begin{pmatrix}
\hat{L}_{11} & \hat{L}_{12} &0 \\
\hat{L}_{21} & \hat{L}_{22} &0   \\
0 & 0 & \hat{L}_{33}  \\
\end{pmatrix}
\begin{pmatrix}
U_{1} \\
U_{2} \\
U_{3} \\
\end{pmatrix}
\equiv
\begin{pmatrix}
-f^{(\bullet)}_1 \\
-f^{(\bullet)}_2 \\
-f^{(\bullet)}_3 \\
\end{pmatrix} \, .
\end{equation}

\subsection{Boundary Conditions for the GPP problem}\label{BCC_GPP}

In this Section we
define appropriate boundary conditions for the GPP problem.
Given the special geometry displayed in Fig. \ref{2D-approximation}, we must distinguish between
those conditions that must be satisfied at the lateral surface $\partial \mathbb{D}_l$
from those at the end surfaces $\partial \mathbb{D}_{\pm}$ of the piezoelectric body.

Although more general settings are possible, we examine here the boundary
conditions corresponding to fixing the tractions and charges at the surfaces
(Neumann-type boundary conditions).

\subsubsection{Lateral surface boundary conditions}

It is assumed here that the problem requires the specification on the
lateral surface of the applied traction force $\bar t_{i}$
and impressed surface charge density $\bar \varrho$.
When working on the 2D cross section
of the problem, this implies the following requirements:
\begin{subequations}\label{EQ:BCS1bis}
\begin{equation}
\hphantom{-}n_{\alpha}\sigma_{\alpha j}= \bar{t}_{j}  \: ,
\end{equation}
\begin{equation}
-n_{\alpha} D_{\alpha}=\bar{\varrho} \: ,
\end{equation}
\end{subequations}
to be satisfied on the boundary $\partial \mathbb{D}^{\mathrm{(2D)}}$.

\subsubsection{End surface boundary conditions}\label{End_BCs}

According to Saint-Venant's principle, originally stated for an elastic problem \citep{iecsan1987saint},
the point-wise specification of the imposed tractions
at the end surfaces of a finite but long body is only necessary if the adjacent regions are to be studied.
Far from those extreme sections, at the central region of the body,
\emph{it is expected}
that the influence of the detailed distribution of end tractions becomes negligible
and the solution of the problem is only affected by the total force and torque \citep{Hwu2010}.
Therefore, to specify completely the GPP problem it should be enough
to prescribe the resulting force $\bm{F}=(F_1,F_2,F_3)$ and torque $\bm{M}=(M_1,M_2,M_3)$, as
well as the net charge $Q$, on the end surfaces:
\begin{subequations}\label{EQ:End_cond}
\begin{equation}
\int_{\mathbb{D}^{\mathrm{(2D)}}}  dx_1 dx_2\,\sigma_{3j}(x_1,x_2)=F_j \: ,
\end{equation}
\begin{equation}
\int_{\mathbb{D}^{\mathrm{(2D)}}}  dx_1 dx_2\,\eta_{j\beta k}\,x_\beta\,\sigma_{3k}(x_1,x_2) =M_j \: ,
\end{equation}
\begin{equation}
-\,\int_{\mathbb{D}^{\mathrm{(2D)}}}  dx_1 dx_2\,D_3(x_1,x_2)=Q \: ,
\end{equation}
\end{subequations}
where $\eta_{jlk}$ is the Levi-Civita tensor.
Note that within the GPP problem the same boundary conditions must apply at
both extreme surfaces, and indeed to every transverse section of the system,
as expressed by \eqref{EQ:End_cond}.

\subsection{Principle of Virtual Work for the GPP Problem}\label{FE}

Finally, we present the simplification of the principle of virtual
work \eqref{Var1} that results from the consideration of the
specific features of the GPP problem.
The structure of the variations compatible with the GPP problem
in combination with the geometry sketched in
Fig.~\ref{2D-approximation} allows, after some algebra,
to decompose the variational problem \eqref{Var1} into two separate
problems.

On one hand, we have the variational
equation in terms of the variations $\delta U_i$ and $\delta \Phi$:
$$
-\int_{\mathbb{D}^{\mathrm{(2D)}}}  dx_1 dx_2\,\left\{ \sigma_{\alpha j} \,\frac{\partial \delta U_j}{\partial x_\alpha}
 +D_\alpha\,\frac{\partial \delta \Phi}{\partial x_\alpha}\right\}
$$
$$
+\int_{\mathbb{D}^{\mathrm{(2D)}}}  dx_1 dx_2\,\left( f^{(\bullet)}_{j}\,\delta U_j-\rho^{(\bullet)}\,\delta\Phi\right)
$$
\begin{equation}\label{Var1GPP}
+ \int_{\partial \mathbb{D}^{\mathrm{(2D)}}} ds \, \bar{t}_j\,\delta U_j-\int_{\partial \mathbb{D}^{\mathrm{(2D)}}} ds \,\bar{\varrho}\,\delta\Phi =0
  \quad ,
\end{equation}
where $ds$ is the counterclockwise length element along $\partial \mathbb{D}^{\mathrm{(2D)}}$.
The enforcement of Eq.~\eqref{Var1GPP} for arbitrary variations leads to the equilibrium equation \eqref{pre-Navier-Poisson} and
the lateral boundary conditions \eqref{EQ:BCS1bis}.

On the other hand, the analysis of the virtual work principle with respect to variations
$\delta \left(\frac{1}{R_1}\right)$, $\delta \left(\frac{1}{R_2}\right)$, $\delta \theta$,
$\delta \varepsilon_{\parallel}$ and $\delta E_\parallel$ gives:
\begin{subequations}\label{fempar}
\begin{alignat}{3}
\left\{\int_{\mathbb{D}^{\mathrm{(2D)}}} dx_1 \,dx_2\,x_1\,\sigma_{33}(x_1,x_2)+M_2\right\}\,\delta \left(\frac{1}{R_1}\right)&=0  \, , \\[1ex]
\left\{\int_{\mathbb{D}^{\mathrm{(2D)}}} dx_1 \,dx_2\,\sigma_{31}(x_1,x_2)-F_1\right\}\,\delta \left(\frac{1}{R_1}\right)&=0  \, , \\[1ex]
\left\{\int_{\mathbb{D}^{\mathrm{(2D)}}} dx_1 \,dx_2\,x_2\,\sigma_{33}(x_1,x_2)-M_1\right\}\,\delta \left(\frac{1}{R_2}\right)&=0\, , \\[1ex]
\left\{\int_{\mathbb{D}^{\mathrm{(2D)}}} dx_1 \,dx_2\,\sigma_{32}(x_1,x_2)-F_2\right\}\,\delta \left(\frac{1}{R_2}\right)&=0 \, , \\[1ex]
\left\{\int_{\mathbb{D}^{\mathrm{(2D)}}} dx_1 \,dx_2\,
\left[x_2\,\sigma_{31}(x_1,x_2)-x_1\,\sigma_{32}(x_1,x_2)\right]
+M_3\right\}\,\delta \theta &=0 \, , \\[1ex]
\left\{\int_{\mathbb{D}^{\mathrm{(2D)}}} dx_1 \,dx_2\,\sigma_{33}(x_1,x_2)-F_3\right\}\,\delta \varepsilon_{\parallel}&=0
\, , \\[1ex]
\left\{-\int_{\mathbb{D}^{\mathrm{(2D)}}} dx_1 \,dx_2\,D_3(x_1,x_2)-Q\right\}\,\delta E_\parallel&=0   \, .
\end{alignat}
\end{subequations}
The enforcement of Eqs.~\eqref{fempar} for arbitrary variations leads to the boundary conditions \eqref{EQ:End_cond}.


\begin{table}[t]
\centering
\caption{Lattice parameters and elastic, piezoelectric and dielectric constants used in the calculations.}
\begin{tabular}{|c c | c| c  | c    | c  | c  | c  |}
\hline
  \multicolumn{2}{|c|}{parameter\quad}
& \multicolumn{1}{c|}{Si}   & \multicolumn{1}{c|}{Ge}
   & \multicolumn{1}{c|}{GaN}  & \multicolumn{1}{c|}{InN}\\ \hline\

  $a_0$     & (\r{A})  & $5.430^{a}$         &  $5.652^{a}$      & $4.50^{b}$          & $4.98^{b}$       \\ \hline
  $C_{11}$  & (GPa)    &$162.0^{c}$          & $128.5^{c}$       &$316.9^{c}$          & $204.1^{c}$     \\
  $C_{12}$  & (GPa)    & $62.8^{c}$          & $45.7^{c}$      & $152.0^{c}$            & $119.4^{c}$          \\

  $C_{44}$  & (GPa)     & $77.2^{c}$          & $66.8^{c}$    & $197.6^{c}$            & $114.1^{c}$       \\ \hline
  $e_{14}$  & (C/m$^2$) &  $-$     &  $-$     & $0.59^{d}$     & $0.84^{d}$       \\\hline
$\epsilon_{11}$  & ($\epsilon_{0}$) & $11.97^{e}$             & $16.00^{e}$      & $9.7^{f}$                & $8.4^{f}$       \\
\hline

\end{tabular}
\caption*{

${}^{a}$reference \citep{Reeber1996}  \\
${}^{b}$reference \citep{vurgaftman2001} \\
${}^{c}$reference \citep{Wang2003}  \\
${}^{d}$reference \citep{Xin2007}  \\
${}^{e}$reference \citep{Madelung2004} \\
${}^{f}$reference \citep{{Hadis2006}}  \\
 }
\label{TBL:parametrosN}
\end{table}

\section{Numerical Results: Application to core-shell nanowires}\label{result}

To illustrate the use and utility of the generalized plane (GPS and GPP) approaches,
we present in this Section numerical results of the strain and electric field distributions
in a
lattice-mismatched core-shell nanowire, a system which has been studied extensively in recent years
\citep{Svensson2008, Morral2008, Pistol2008, Montazeri2010, Ben-Ishai2010, Wong2011, Ferrand2014}.

In this work we have opted to solve numerically for the strain and electric fields by using
the finite element method (FEM),
as implemented in the software platform \textit{COMSOL Multiphysics} \citep{COMSOL2010}.
However, the 2D modules of this platform only provide the plane piezoelectric approximation.
Therefore, we have  adapted the software by incorporating variational equations \eqref{Var1GPP} and
\eqref{fempar} to include the solution for the warping field $U_3(x_1,x_2)$ and the parameters
$\left(\varepsilon_{\parallel}, \frac{1}{R_1}, \frac{1}{R_2},  \theta,   E_\parallel\right)$
\citep{aMengistu14,bMengistu14}.
We will first study the purely elastic problem as manifested in a Ge/Si(111) core-shell nanowire \citep{Lauhon2002, Musin2005, Goldthorpe2008, Ben-Ishai2010}. In a second example,  where we want to consider a system that exhibits reasonably strong piezoelectric effects, we have chosen a zincblende InN/GaN(111) core-shell nanowire, which has also been studied recently \citep{Kim2009, Cui2012, Sangeetha2013, Wu2014, Tchernycheva2014}.

We have performed 2D generalized plane calculations according to the framework introduced in Sec. \ref{GPP-problem}
for a nanowire of radius $R_\mathrm{NW}$ (to be
precisely specified below). In addition, in order to test the quality of our GPS/GPP approaches,
in both cases we have also performed
fully 3D computations for a finite but long nanowire.
The nanowire length $L$ is chosen
such that the system has a
high aspect ratio, i.e., $L/(2R_\mathrm{NW})\gg 1$. The later calculations (more precisely, the results at the central cross section) are then compared with those obtained from the 2D GPS/GPP approaches. From the comparison, we can ascertain under what circumstances the generalized plane approaches represent a good approximation to the central part of 3D systems, thereby quantifying their accuracy and limitations.

\begin{figure}[!ht]
\centering
\includegraphics[width=1\textwidth]{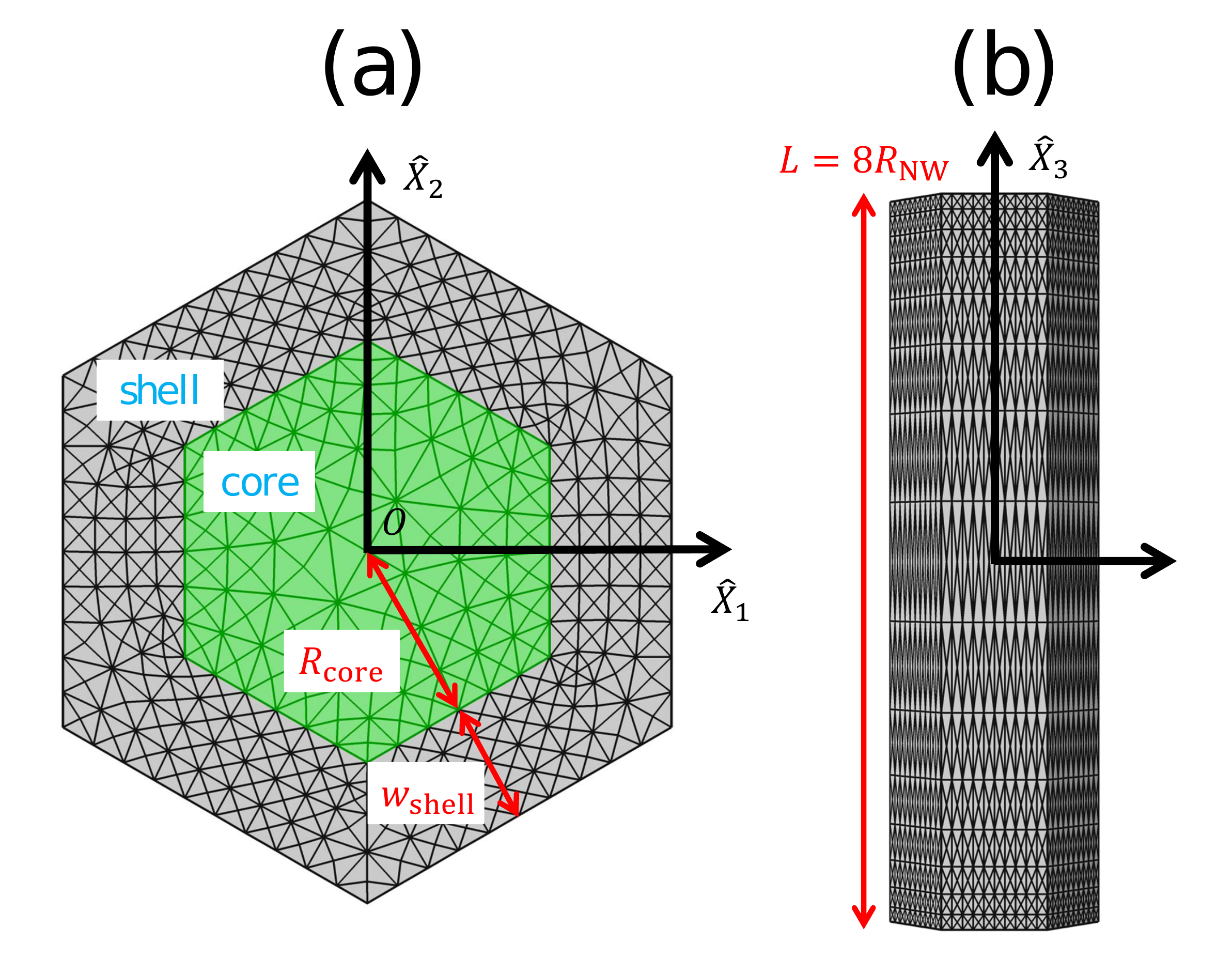}
\caption{Geometry of the core-shell
nanowire investigated,  with the
meshes employed in the FEM calculations. (a) The hexagonal cross section is characterized by the values of the core radius $R_\mathrm{core}$
and the width of the shell $w_\mathrm{shell}$. The total radius of the nanowire is then $R_\mathrm{NW}=R_\mathrm{core}+w_\mathrm{shell}$. (b) Lateral view of the finite nanowire considered for the 3D
simulations. The length used is $L=8R_\mathrm{NW}$.}
\label{nanowire_scheme}
\end{figure}

\subsection{Non-piezoelectric problem in a core-shell nanowire}\label{Elastic_problem}

First, we will apply the GPS approach to study a purely elastic problem corresponding to a
Ge/Si(111) core-shell nanowire. The core is made of Ge and the shell is made of Si. The geometry of the nanowire is shown in Fig.~\ref{nanowire_scheme}. The cross section is assumed to be hexagonal and it is characterized by the values of the core radius $R_\mathrm{core}$ and the width of the shell $w_\mathrm{shell}$. The total radius of the nanowire is then $R_\mathrm{NW}=R_\mathrm{core}+w_\mathrm{shell}$. Within the linear elastic continuum theory the distribution of the strain fields does not depend on the absolute dimensions but on their relative size, in this case on $w_\mathrm{shell}/R_\mathrm{core}$.

Ge and Si have diamond crystalline structure and therefore exhibit macroscopic cubic $O_h$ symmetry, which is fully
taken into account in the following calculations. We have chosen the longitudinal axis $\hat{X}_3\equiv Z$ of the nanowire to be along the [111] crystallographic direction, whereas the axes $\hat{X}_1\equiv X$ and $\hat{X}_2\equiv Y$ are taken along $[1 0 \bar{1}]$ and $[\bar{1}2\bar{1}]$ directions, respectively. The matrix of elastic constants $\hat{C}_{IK}$ corresponding to the above axes can be found in Appendix~\ref{APP:Appendix A}. It is apparent there that, since $\hat{C}_{14},\hat{C}_{24}(=-\hat{C}_{14}) \neq 0$,
it is not possible the decoupling leading to
Eq. \ref{EQ:GPP_MATRIX4}, and it is expected a nonvanishing warping function $U_3$ coupled to the in-plane deformation $(U_1, U_2)$.

We assume that the nanowire is free from external traction and body forces (i.e., $\bar{t}_j,F_j,M_j,
f_j=0$),
so that the strain is solely induced by the internal lattice mismatch
between the core (inclusion, $I$) and shell (matrix, $M$) regions, through the body force $f_i^{(0)}$ given by Eq.~\eqref{EQ_INCLUSION:Equ1}. The lattice parameters and elastic constants used in the calculations can be found in Table~\ref{TBL:parametrosN}. For cubic materials there is only one lattice parameter, $a_i \to a_0$, and therefore the misfit strain is diagonal, with magnitude $\varepsilon^{(\mathrm{misfit})}=-0.039$ (3.9\%). The negative sign indicates that
the strain is compressive.

\begin{figure}[!ht]
\centering
\includegraphics[width=\textwidth]{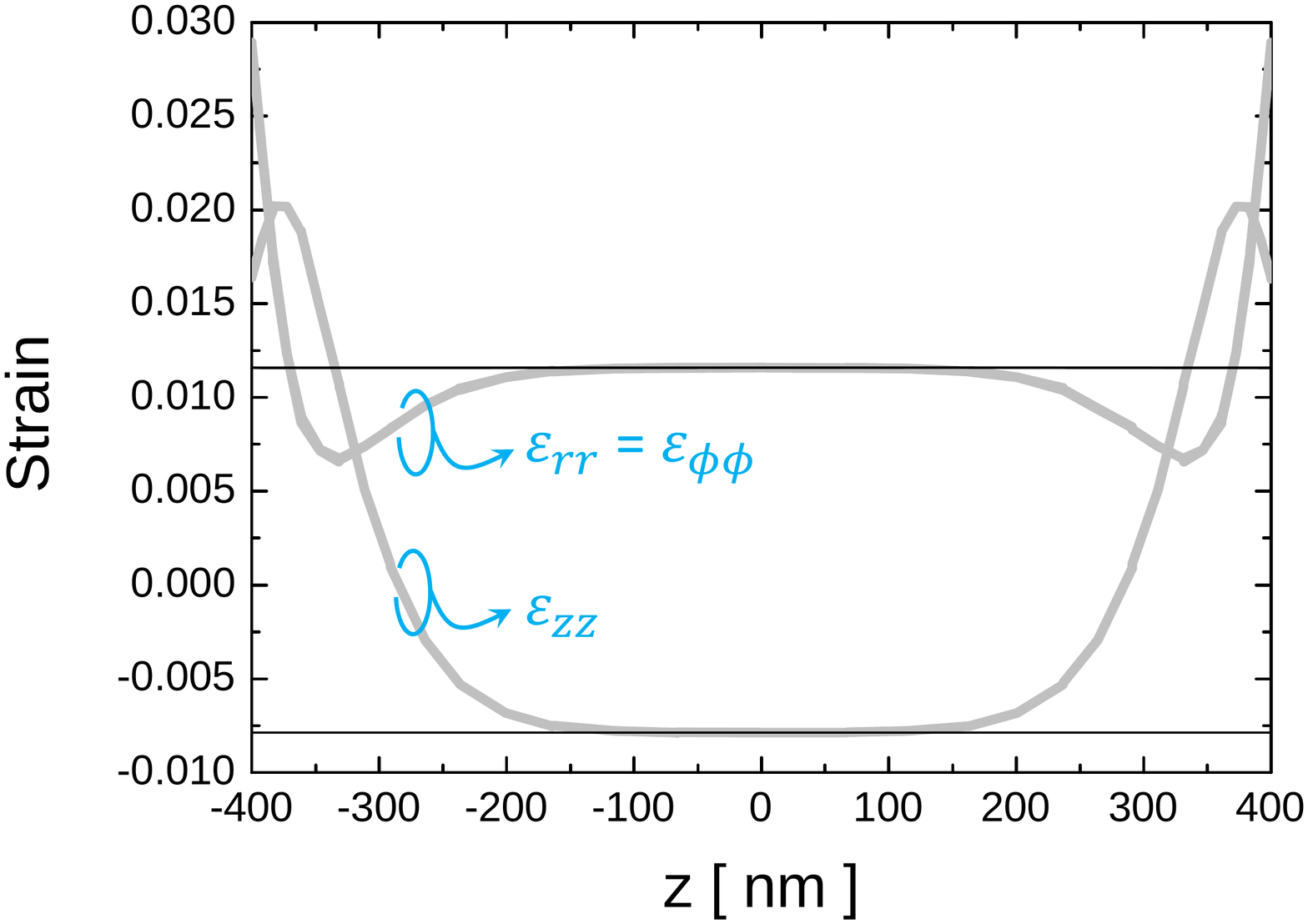}
\caption{The gray thick curves represent the linescans of the relaxation strain components  $\varepsilon_{rr}$, $\varepsilon_{\phi\phi}$ and $\varepsilon_{zz}$ along the $Z$ axis of the finite nanowire of Fig.~\ref{nanowire_scheme}(b), as obtained from the 3D calculations. Note that $\varepsilon_{rr}=\varepsilon_{\phi\phi}$ along the nanowire axis. For comparison, the results corresponding to an infinite nanowire as obtained by means of the GPS approach are also displayed as black horizontal lines.}
\label{Si-Ge_axis_comparison}
\end{figure}

\begin{figure}[!ht]
\centering
\includegraphics[width=1\textwidth]{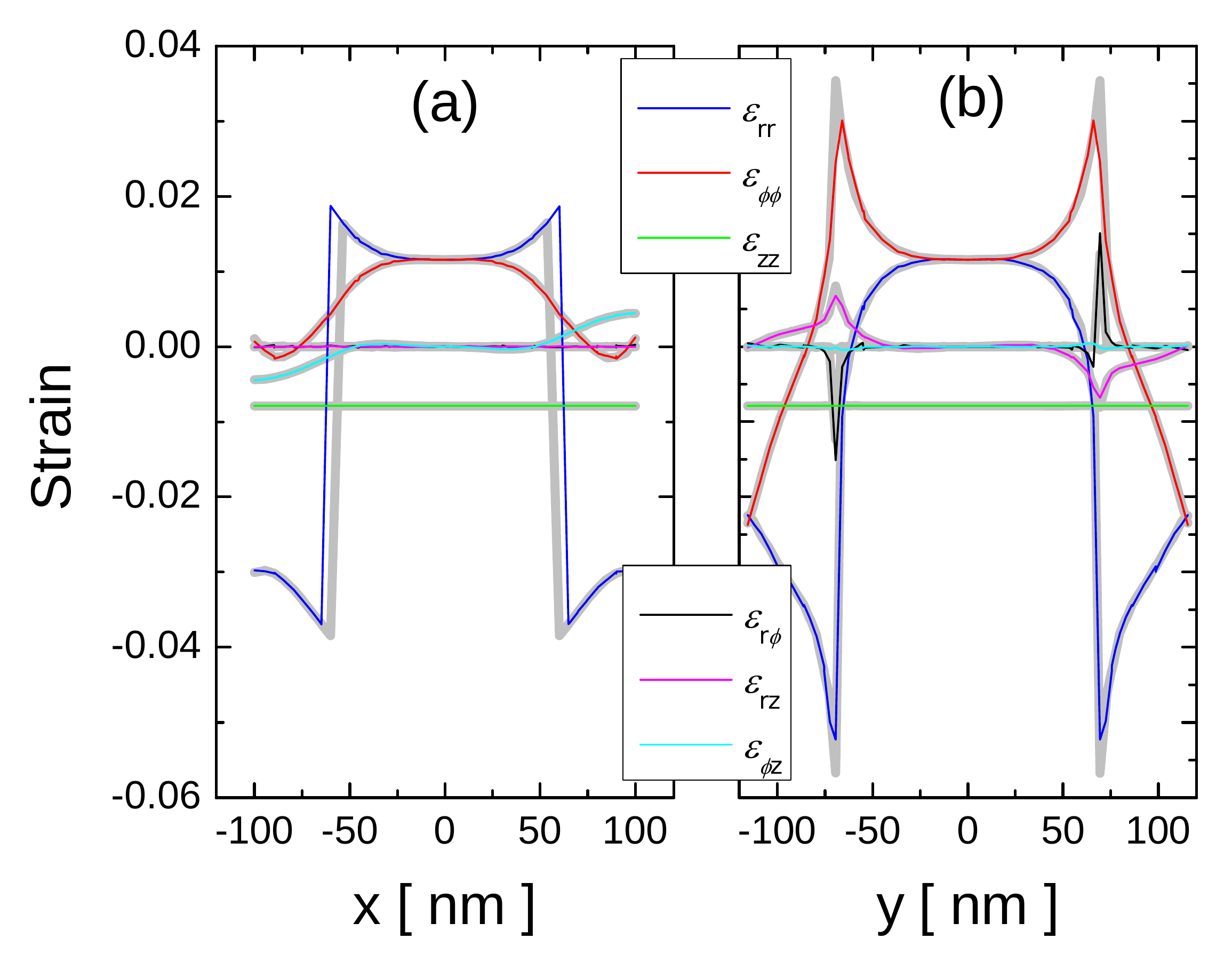}
\caption{Linescans of the
relaxation strain components in the nanowire transverse section, (a) along the $X$ axis  and (b) along the $Y$ axis,
as obtained with the 3D calculations at the central cross section $z=0$ of the finite nanowire (thick gray lines) and as obtained by means of the 2D GPS calculations (thin coloured lines).}
\label{Si-Ge_plane_comparison}
\end{figure}

We have taken the following numerical values for the geometry parameters (see Fig.~\ref{nanowire_scheme}): $R_\mathrm{core}=60$ nm,
$w_\mathrm{shell}=40$ nm, so that $R_\mathrm{NW}=R_\mathrm{core}+w_\mathrm{shell}=100$ nm. Therefore, we have $w_\mathrm{shell}/R_\mathrm{core}=2/3$. For the 3D calculations we have taken $L=8R_\mathrm{NW}$, so that the aspect ratio is $L/(2R_\mathrm{NW})=4$, which will be shown
to represent well the limiting case $L/(2R_\mathrm{NW})\gg 1$. Note that the output of our numerical calculations
is the displacement associated to the relaxation strain $\varepsilon_{ij}$ with respect to the reference lattice
$a_0^{(\mathrm{ref})}$ (see discussion in Section \ref{INC}). In the numerical calculations below we have taken as reference lattice that specified by $a_0^{(\mathrm{ref})}=(a_0^{(M)}+a_0^{(I)})/2$.
In the 3D-2D comparisons of Figs.~\ref{Si-Ge_axis_comparison} and \ref{Si-Ge_plane_comparison}
it has been represented the relaxation strain $\varepsilon_{ij}$ because it is the direct output from the FEM calculations.
We note that, although the concrete numerical values obtained for $\varepsilon_{ij}$ depend on the choice of $a_0^{(\mathrm{ref})}$, the relevant total local strain $\varepsilon^{(T)}_{ij}$ can be recovered eventually by adding the strain associated to the reference lattice $\varepsilon^{(0)}_{ij}$, as shown in Eq.~\eqref{Total_strain}.

The strain components to be analyzed below are referred
to the system of axes $O\,\hat{X}_1 \hat{X}_2 \hat{X}_3$ presented in Fig.~\ref{nanowire_scheme},
although for the discussion below it
is preferable to express them in cylindrical coordinates $(r,\phi,z)$ rather than in Cartesian ones
(see Appendix~\ref{APP:Cylindrical_components}).

In the first place,
we display in Fig.~\ref{Si-Ge_axis_comparison} the linescans of the the relaxation strain components $\varepsilon_{rr}$, $\varepsilon_{\phi\phi}$ and $\varepsilon_{zz}$ along the $Z$ axis of the finite nanowire, as obtained with the 3D calculations. For comparison, the values obtained by means of the
2D GPS calculations at the center of the NW cross section are also indicated as horizontal lines. The GPS approach gives an axial strain $\varepsilon_\parallel=-0.00787$ and no bending (the calculated maximum bending strains are $|R_\mathrm{NW}/R_1|, |R_\mathrm{NW}/R_2| < 10^{-8}$, which can be considered zero within the numerical error),
so that $\varepsilon_{zz}=\varepsilon_\parallel$. The calculated maximum torsion  strain $(\theta R_\mathrm{NW})/2=2.3\times 10^{-8}$ is also  zero  within the numerical error.
The absence of bending   in this particular case is due to the concentric nature
of the core-shell system. We have checked (both in 3D and 2D calculations)
that a nonvanishing bending
of the nanowire would appear as soon as the symmetry of the cross section is broken, e.g, by imposing
that the geometric centers of the core and shell regions do not coincide.
The first conclusion we want to draw from Fig.~\ref{Si-Ge_axis_comparison} is that the 3D calculation for the long nanowire studied here presents two well
differentiated regions. There is the region $|z|\leq 150$ nm  around the center of the nanowire
(\,otherwise stated, at distances away from the end surfaces larger than $250 \: \mathrm{nm} = 1.25(2R_\mathrm{NW})$\,),
where the strain is essentially
independent of $z$,
with  no  strain component deviating by more than 4\% from the central values at $z=0$.
On the contrary, in the region within a distance of $\sim 1.25(2R_\mathrm{NW})$ from the end surfaces
the strains are rather inhomogeneous.
Moreover,
the agreement between the results for the central cross section of the finite model and
those of an infinite nanowire is better than 99.8\%.
The above picture already indicates that our finite nanowire has sufficient length so that the strain field at
its central portion corresponds to that of the infinite model modeled by our GPS approach, thus giving a numerical confirmation of Saint-Venant's principle.
In Fig.~\ref{Si-Ge_plane_comparison}, scans of the various relaxation strain components are shown now along two
different directions on the nanowire cross section. The linescans for the strain calculated by means of the GPS approach show again an excellent agreement with the 3D results at the central cross section of the finite wire.
The linescans along the $Y$ axis show that strain singularities appear at
the bimaterial corners. A more careful consideration would be needed if one were interested in a detailed
analysis of the behavior around these singularities, since they may strongly affect the accuracy of the FEM
results. This problem can be alleviated, e.g., by applying strong h- or p-refinements of the mesh in the neighbourhood of such
corners. Such refinements have not been incorporated in the simulations presented in this paper, but we note that it would
be much easier to capture the strain singularities within GPP modeling than in the standard 3D FEM
analysis.

\begin{figure}[!ht]
\centering
\includegraphics[width=1\textwidth]{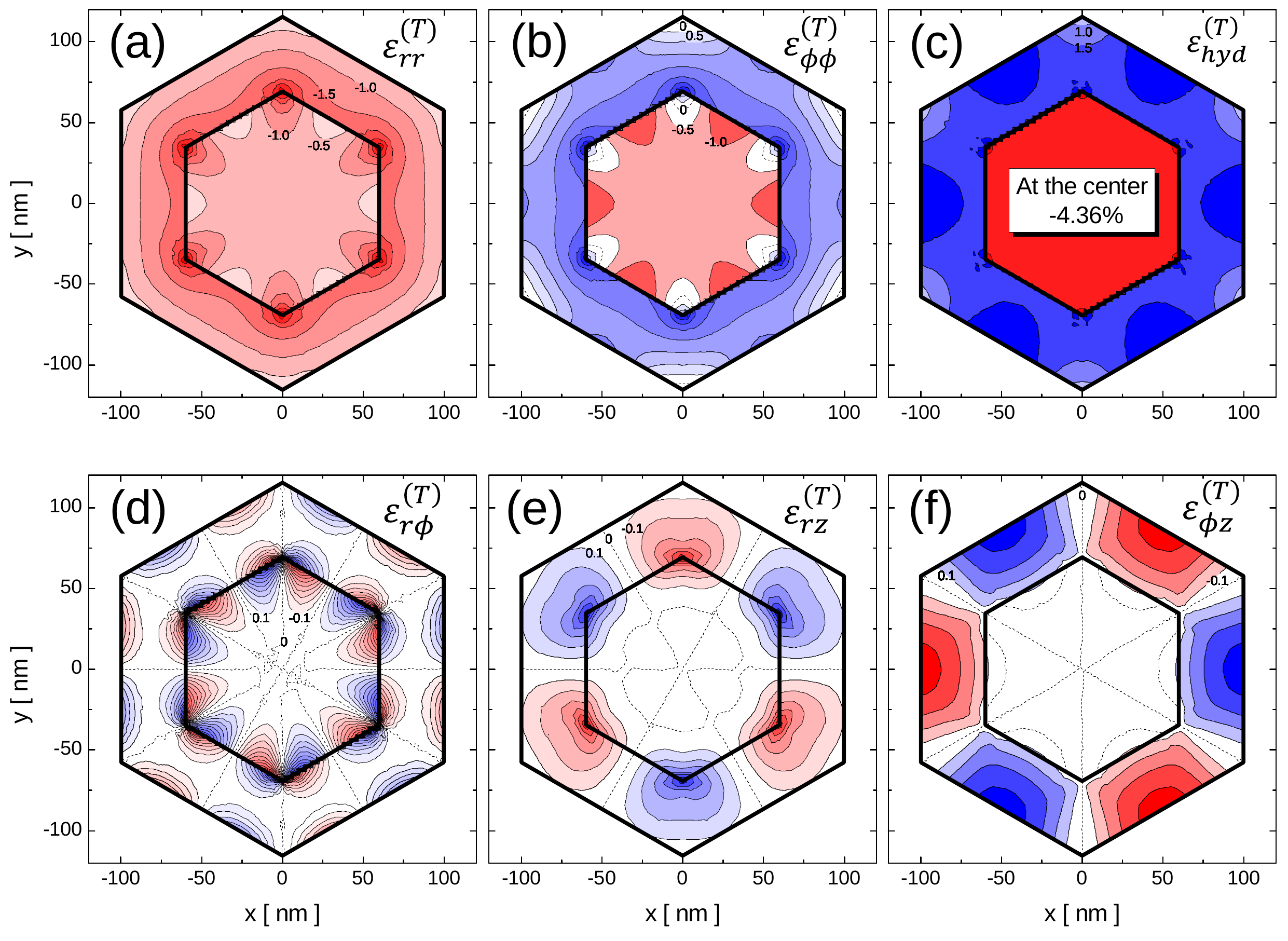}
\caption{In-plane distribution of the total strain
in an infinite core-shell nanowire as calculated by the GPS approach. Panels (a)-(c) show the strain components $\varepsilon^{(T)}_{rr}$, $\varepsilon^{(T)}_{\phi\phi}$, $\varepsilon^{(T)}_\mathrm{hydro}=\varepsilon^{(T)}_{rr}+\varepsilon^{(T)}_{\phi\phi}+\varepsilon^{(T)}_{zz}$
(contour lines are displayed for strain values in steps of 0.5$\%$; the zero strain contour is drawn with a dashed line). Panels (d)-(f) show the shear strains $\varepsilon^{(T)}_{r \phi}$, $\varepsilon^{(T)}_{rz}$  and $\varepsilon^{(T)}_{\phi z}$ (here contour lines are displayed for strain values in steps of 0.1$\%$).}
\label{Si-Ge_plane_2D}
\end{figure}

Since the above analysis confirms the 2D GPS approach as a very reliable tool to model long nanowires, we use it
in the following for a detailed description and understanding of the
strain distribution in the nanowire cross section.
In Fig.~\ref{Si-Ge_plane_2D} we show the contour plots of the total strain components $\varepsilon^{(T)}_{ij}$ in the $XY$ plane as provided by the GPS approach.
We will try to elucidate the obtained anisotropic strain distribution with the following comments, that
show that the overall features can be traced back to the initial lattice mismatch and the geometry.

\begin{itemize}

\item The axial strain component $\varepsilon^{(T)}_{zz}$ (not shown in the figure)
exhibits a simple step-like
profile, whose values can be obtained easily from the calculated $\varepsilon_\parallel$ and the initial strain $\varepsilon^{(0)}_{ij}$:
a compressive strain in the core
$\varepsilon^{(T)}_{zz}(\mathrm{core})=-0.0275$
and tensile strain in the shell
$\varepsilon^{(T)}_{zz}(\mathrm{shell})=+0.0126$ are so obtained.
These values are consistent with the lattice mismatch. The core material (Ge) has a larger lattice constant than the shell material (Si),
and the energetically most favorable configuration in the deformed core-shell system corresponds to a common axial lattice constant of value $a_z=5.497$ {\AA} in between those of bulk Ge and bulk Si .

\item Now we focus on the
core-shell interface. If one looks in particular at the
central part of that interface, one sees a
difference in sign between the values of the tangential component $\varepsilon_{\phi \phi}$ at the core and shell sides of the interface,
the shell material being expanded
($\varepsilon_{\phi \phi}(\mathrm{shell})>0$) whereas the core material is compressed
($\varepsilon_{\phi \phi}(\mathrm{core})<0$), much in the same way as $\varepsilon_{zz}$, and for the same reasons. Concerning the radial strain $\varepsilon_{r r}$ in the shell material, the compressive character of $\varepsilon_{zz}$ and $\varepsilon_{\phi \phi}$, and the freedom at the outer surface of the shell
determines that $\varepsilon_{r r}(\mathrm{shell})<0$, i.e., that the shell is radially compressed. Although the same rationale would imply that the $\varepsilon_{r r}$ of the core would tend to be positive,
the fact that the core is constrained by the shell makes that response
impossible to attain and the radial strain is eventually slightly compressive, $\varepsilon_{r r}(\mathrm{shell})<0$.

\item The measure of the local volume deformation, the hydrostatic strain
$\varepsilon^{(T)}_\mathrm{hydro}=\varepsilon^{(T)}_{rr}+\varepsilon^{(T)}_{\phi\phi}+\varepsilon^{(T)}_{zz}$, is depicted in Fig.~\ref{Si-Ge_plane_2D}(c).
In contrast to the individual strain components, which have a complicated inhomogeneous space distribution, the hydrostatic strain has a very simple behavior:
The volume of the shell (core) is expanded (compressed) in almost an uniform manner.

\item In general, the values of the strain components change smoothly
when moving from the core-shell interface towards the nanowire center (in the core) and  towards the surface (in the shell). However, when moving towards the corners they experience a stronger variation. This behavior can also be seen in the linescans of Fig.\ref{Si-Ge_plane_comparison}(b). The corner geometry allows for some partial stress relief, as manifested by the increased importance of the shear strain $\varepsilon_{r\phi}$.

\item
The small but nonvanishing values for the out-of-plane shear strain components $\varepsilon_{rz}$ and $\varepsilon_{\phi z}$ displayed in Figs.~\ref{Si-Ge_plane_2D}(e) and (f) are a direct consequence of $U_3(x_1,x_2)\neq 0$,
and correspond to the fact that a cross section in the $XY$ plane is warped into a non-flat surface in the strained
wire (warping effect). In other words, when the core pushes outwards on the shell, it is energetically favorable for the system to respond by
not only becoming deformed in the plane, but also by warping out of the $XY$ plane. One important aspect of our 2D GPS calculation is that it is able to fully capture that warping effect, which is impossible to obtain by working under
the standard plane strain approximation that forces the warping function $U_3$ to vanish.

\end{itemize}

\subsection{Piezoelectric problem in a core-shell nanowire}\label{Piezoelectric_problem}

Next, we will present the GPP results for a fully-coupled piezoelectric problem.
The problem corresponds to a zincblende core-shell nanowire oriented  along
the [111] direction. The core is made of InN and the shell is made of GaN.
The geometry of the nanowire is the same shown in Fig.~\ref{nanowire_scheme}.
The elastic, piezoelectric and dielectric tensors of zincblende materials
exhibit cubic $T_d$ symmetry, which is fully taken into account in the calculations.
The Voigt matrices in the reference frame associated to the nanowire, $\hat{C}_{IK}$, $\hat{e}_{nI}$ and
$\hat{\varepsilon}_{mn}$, can be found in Appendix \ref{APP:Appendix A}.
Now, besides the nonvanishing elastic constants
$\hat{C}_{14},\hat{C}_{24}(=-\hat{C}_{14}) \neq 0$, there appear nonvanishing
piezoelectric constants $\hat{e}_{15},\hat{e}_{24}(=\hat{e}_{15})\neq 0$,
so the system will  exhibit a nonvanishing warping function $U_3$, coupled to both $(U_1,U_2)$ and  $\Phi$.

We assume again that the nanowire is free from external traction and body forces
(i.e., $\bar{t}_j,F_j,M_j,
f_j=0$), as well as from external charges (i.e., $\bar{\varrho},Q,\rho=0$),
so that the only cause for the deformation and potential fields is the
lattice mismatch, through the body force $f_i^{(0)}$ and the charge density $\rho^{(0)}$
given in Eqs.~\eqref{INCLUSION}. The lattice parameters and material constants used can be found in
Table~\ref{TBL:parametrosN}. In this case, the diagonal misfit strain between the core (inclusion, $I$) and shell (matrix, $M$) is $\varepsilon^{(\mathrm{misfit})}=-0.0964$ (9.64\%).
For the following calculations we have taken the same cross section geometry and length as in Sec. \ref{Elastic_problem}. As before, the relaxation strain is calculated starting from a reference lattice with $a_0^{(\mathrm{ref})}=(a_0^{(M)}+a_0^{(I)})/2$, and the strain and electric fields will be expressed in cylindrical coordinates  (see
Appendix~\ref{APP:Cylindrical_components}).

\begin{figure}[!ht]
\centering
\includegraphics[width=1\textwidth]{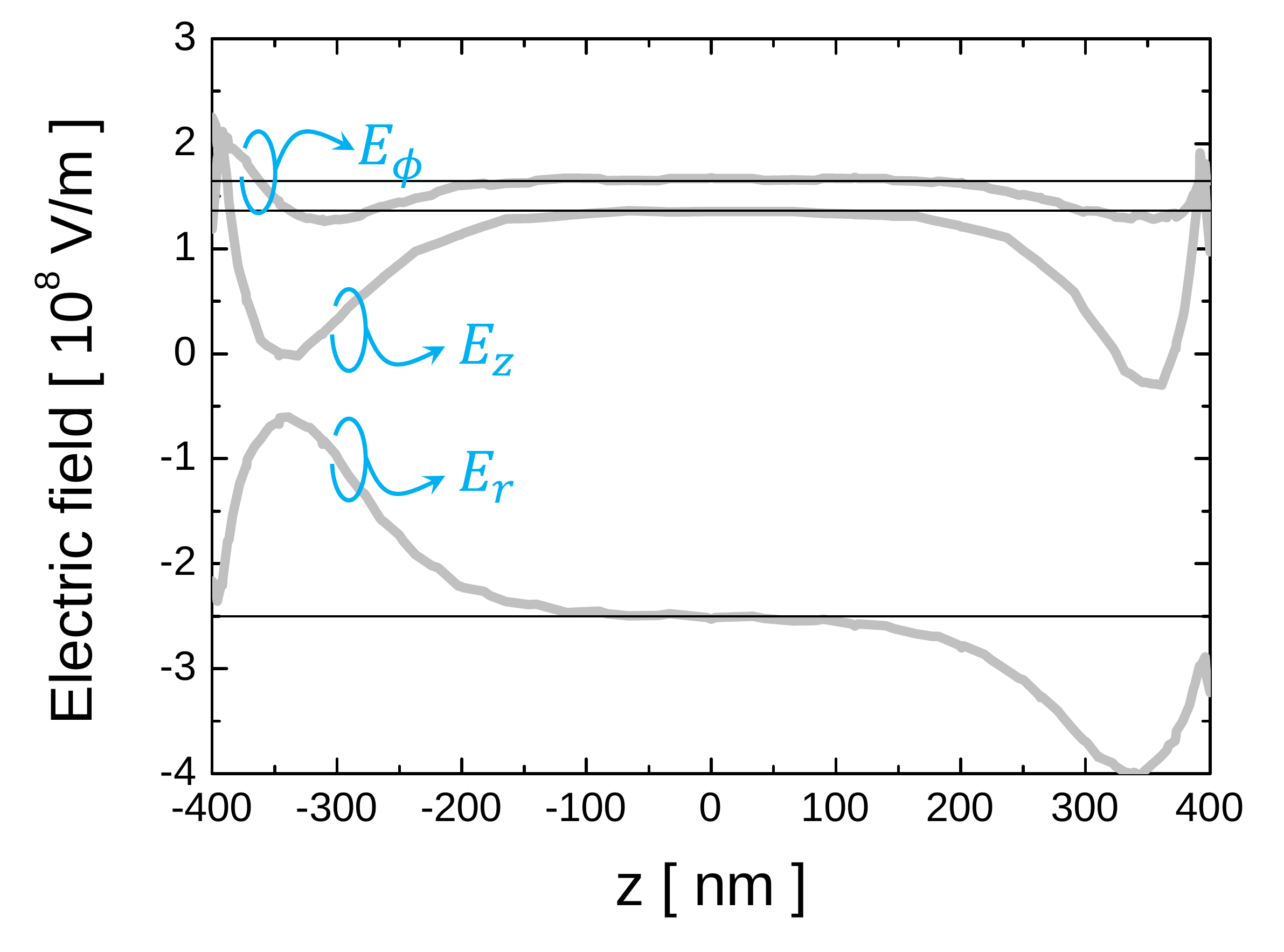}
\caption{The gray thick curves represent the linescans of the electric field components $E_r$, $E_\phi$ and $E_z$ along the longitudinal axis of the finite nanowire passing through $(x=0.5 R_{\mathrm NW},y=0.5 R_{\mathrm NW})$, as obtained from the 3D calculations. For comparison, the results corresponding to an infinite nanowire as obtained by means of the GPP approach are also displayed as black horizontal lines.}
\label{In-GaN_axis_comparison_field}
\end{figure}

\begin{figure}[!ht]
\centering
\includegraphics[width=1\textwidth]{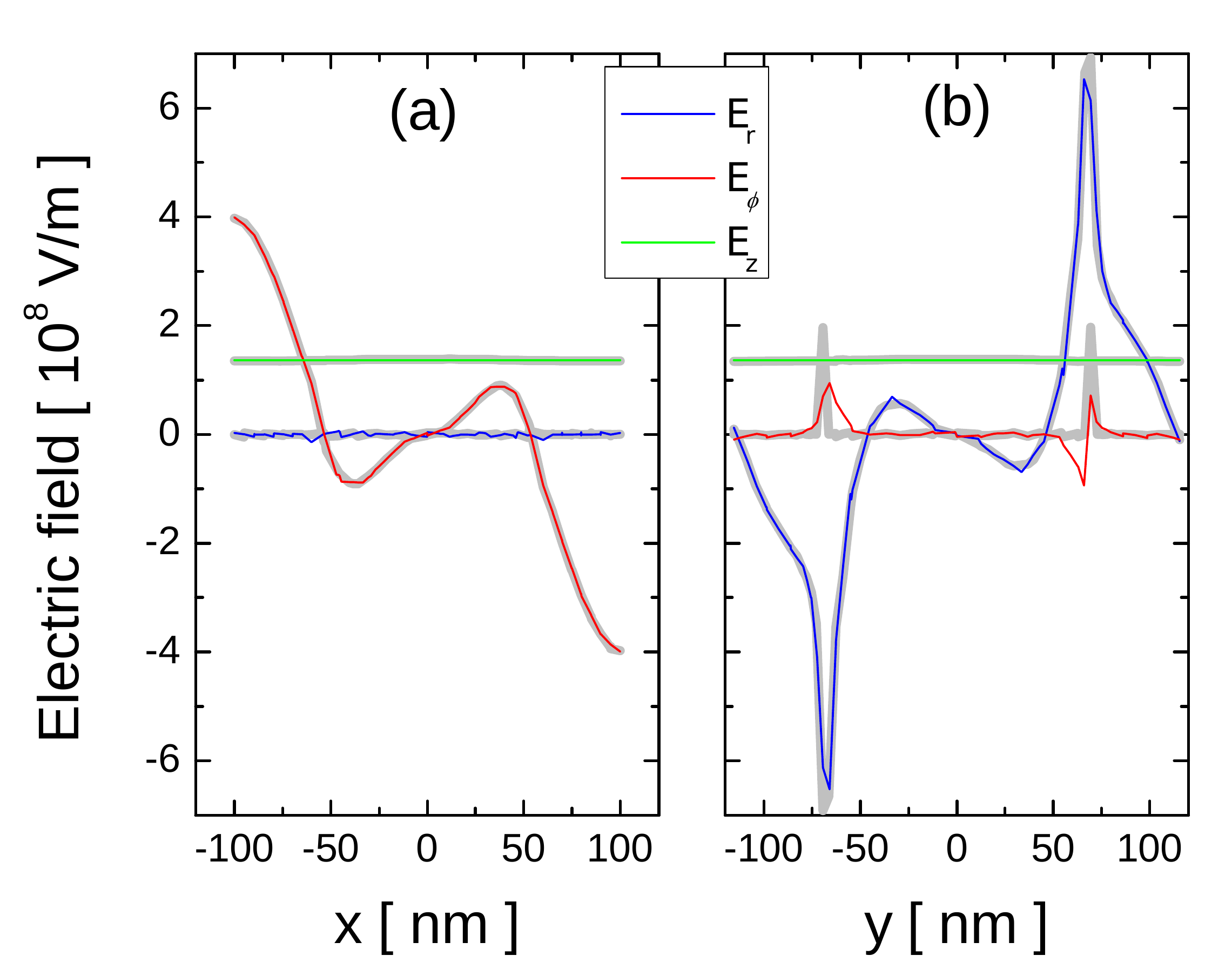}
\caption{Linescans of the field components in the nanowire transverse section, (a) along the $X$ axis and (b) along the $Y$ axis, as obtained with the 3D calculations at the central cross section
$z = 0$ of the finite nanowire (thick gray lines) and as obtained by means of the 2D GPP
calculations (thin coloured lines).}
\label{InN-GaN_plane_comparison_field}
\end{figure}

Since the the sign of the misfit strain is the same, the strain fields in this case follow a pattern qualitatively similar to that for the non-piezoelectric nanowire in
Sec.~\ref{Elastic_problem}. Consequently, we focus here on the piezoelectric field and potential. In the first place, we will analyze the linescans of the field along a longitudinal axis of the finite nanowire. Due to symmetry reasons, $E_r, E_{\phi}=0$ along the $Z$ axis, so we have chosen to display in Fig. \ref{In-GaN_axis_comparison_field} a linescan of the field components, as obtained from the full 3D calculation, along a longitudinal off-center axis crossing the transverse section through coordinates $(x=0.5 R_{\mathrm NW},y=0.5 R_{\mathrm NW})$. For comparison, the corresponding values obtained
by means of the GPP approach are also indicated as horizontal lines. In particular, the GPP approach gives an axial field  $E_z= E_\parallel=1.36\times 10^8$ V/m, which is uniform throughout the transverse section.
As for the strain, we distinguish also in the field profile two regions.
In the central region, for distances away from the end surfaces larger
than $1.25(2R_{\mathrm NW})$  (i.e., $|z|<150$ nm ), the field is rather uniform along the axis (e.g., $E_z$ does not deviate by more than 5\% from the value at $z=0$). On the other hand,  within a distance of $1.25(2R_{\mathrm NW})$ from
the end surfaces, the field varies considerably, mainly as a consequence of the over relaxation effect in the strain components. The results for the central cross section of the finite model are again very well approximated by those of an infinite nanowire as calculated by means of the 2D GPP approach (the agreement at $z=0$ being better than 99.3\% for $E_z$ and better than 98\%   for $E_r$ and $E_\phi$).
These results represent a numerical confirmation that Saint-Venant's principle works well
also for the fully-coupled piezoelectric nanowire problem.
In Fig.~\ref{InN-GaN_plane_comparison_field}
the various field components are shown along two different directions on
the nanowire cross section. The field linescans
calculated by means of the GPP approach show again a remarkable agreement
with the 3D results at the central cross section of the finite wire,
thus confirming the reliability of the GPP approach to simulate the central region of high aspect-ratio piezoelectric problems.
Figure~\ref{InN-GaN_plane_comparison_field}(b) evidences that $E_r$ has a singular behavior at the bimaterial corners, which, as said before, is difficult to capture accurately with the FEM calculations unless specific procedures are adopted.
This inaccuracy of our FEM results is also manifested in the spurious finite values of $E_\phi$ at the corners,
that should be exactly zero by symmetry arguments.

\begin{figure}[!ht]
\centering
\includegraphics[width=1\textwidth]{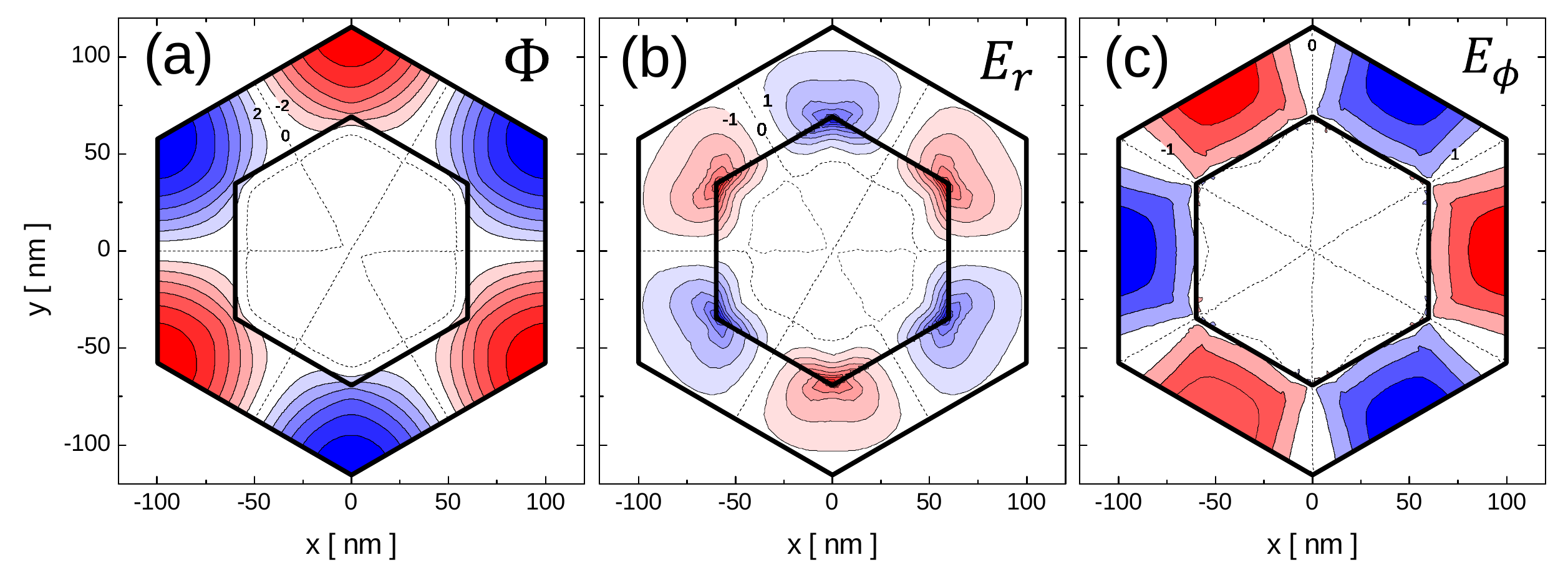}
\caption{
In-plane distribution of: (a) the potential $\Phi$ (potential contour lines
in steps of 2 V), and field components (b) $E_r$ and (c) $E_\phi$
(field contour lines in steps of $10^8$ V/m),
in an infinite core-shell nanowire, as calculated by the GPP approach.
}
\label{InN-GaN_plane_2D_potential_field}
\end{figure}

Finally, we illustrate in Fig.~\ref{InN-GaN_plane_2D_potential_field} the distribution of the in-plane piezoelectric potential profile $\Phi(x_1,x_2)$ and electric field components across the $XY$ plane for an infinite wire modeled using the GPP model. Remember that $E_z$ is uniform and equal to $1.36\times 10^8$ V/m.
It is apparent in Fig.~\ref{InN-GaN_plane_2D_potential_field}(a) that
the highest/lowest value of piezoelectric potential ($\pm 13.6 V$) locate in an alternated manner at the external corners of the GaN shell, while the InN core is mostly at zero potential. The associated in-plane field distribution is shown in Figs. \ref{InN-GaN_plane_2D_potential_field}(b)  and \ref{InN-GaN_plane_2D_potential_field}(c). The maximum values of the radial component of the in-plane field $E_{r,\mathrm{max}} = 8.95 \times 10^8$ V/m are confined at the corners of the core-shell interface. On the other hand, $E_{\phi, \mathrm{max}} = 3.97 \times 10^8$  V/m is located at the outer surface of the shell, between the corners.


\section{Summary and Conclusions}\label{Summary and conclusions}

In this work, we have introduced and developed systematically a 2D methodology for the solution of a certain
class of fully-coupled piezoelectric problems: The theoretical framework defined by the set of equations Eq.~\eqref{EQ:displacements}-~\eqref{EQ:Body_force},
together with boundary conditions \eqref{EQ:BCS1bis} and \eqref{EQ:End_cond}, constitutes the mathematically
2D \emph{generalized plane piezoelectric (GPP)} problem.
The GPP problem represents a \emph{very good approximation}
for the central region of a 3D \emph{finite (length $L$) but slender system}
whose transverse section, material properties, loads (forces and charges) and boundary (and interface) conditions
are translationally invariant along its longitudinal direction.
Alternatively, if the limit $L\to \infty$ is
taken, the above approximation becomes an exact picture for the whole system, and therefore the GPP can also be viewed as an \emph{exact representation} for an \emph{idealized infinite system}.
The first obvious advantage of the presented procedure is that, being a 2D approach,
it is cheaper computationally than the original 3D treatment.
Moreover,
the GPP approach is able to accommodate any geometric cross section, crystal class symmetry, axis orientation
and a wide range of compatible boundary conditions, corresponding to different kinds of externally applied stresses (such as hydrostatic pressure, bending moments...) and imposed surface charges.
For instance, when combined with the Eshelby methodology for a coherent piezoelectric inclusion problem, the GPP approach is well suited to handle piezoelectric problems in elongated lattice-mismatched heterostructures.
These possibilities have been illustrated by the
numerical simulation, based on a FEM implementation of the GPP approach, of indefinite lattice-mismatched core-shell nanowires along [111] direction
made of diamond Ge/Si and zincblende piezoelectric InN/GaN materials. Corresponding 3D simulations have also
been performed on finite but long versions of those systems. The 3D-2D comparisons show that for these systems, the behavior of the 3D solutions  (strain and electric fields) at distance $\gtrsim 1.25D$ (where $D$ is the largest dimension of the cross section) from the end surfaces  is very well approximated by the predictions of the 2D GPP approach, in both non-piezoelectric and piezoelectric problems.
This serves as a numerical illustration of the validity of Saint-Venant's
principle in the case of 3D fully-coupled piezoelectric problems.
Note however that the precise scale at which this approximation works depends on the material properties.
The superiority of the GPP approach
is clearly manifested in the core-shell nanowire simulations where warping effects, completely absent in the standard 2D plane piezoelectric approximation, are perfectly taken into account. In conclusion, the GPP approach provides a versatile procedure to study accurately and with moderate computing resources the details of the strain and electric field distribution in elongated piezoelectric systems.


\section*{Acknowledgements}

This  work  has  been financially  supported  by  the  European  Union  through  the  Grant Agreement No.265073-NANOWIRING of the Seventh Framework Program, and by the Ministry of Finances and Competitiveness (MINECO) of Spain through Grants CSD2010-00044 of the Programme ''Consolider Ingenio 2010'' and MAT2012-33483.
The authors thankfully acknowledge the computer resources, technical expertise and
assistance provided by the "Centre de C\`{a}lcul de la Universitat de Val\`{e}ncia" through the
use of Tirant, the local node of the Spanish Supercomputation Network.

\renewcommand{\theequation}{A-\arabic{equation}}
  \setcounter{equation}{0}  

\newpage

\begin{appendices}
\section{Material tensors of the cubic system in the Voigt notation}\label{APP:Appendix A}

In the Voigt notation, the elastic stiffness (fourth rank) tensor $C_{ijkl}$,
the piezoelectric (third-rank) tensor $e_{nij}$, and the dielectric (second-rank) tensor $\epsilon_{mn}$ are conveniently represented by matrices $C_{IK}$, $e_{nI}$ and $\epsilon_{mn}$ ($I,K=1,\dots,6$, $m,n=1,2,3$) \citep{Nye1985}. For crystalline materials belonging to the cubic system (crystal classes $T$ and $T_d$), when referred to the crystallographic axes
($X_1 \parallel [100]$, $X_2 \parallel [010]$, $X_3 \parallel [001]$), these matrices are given by \citep{Nye1985}:
\begin{equation}
C_{IK}\leftrightarrow
\begin{pmatrix}
C_{11} & C_{12} & C_{12} & 0 & 0 & 0 \\
C_{12} & C_{11} & C_{12} & 0 & 0 & 0 \\
C_{12} & C_{12} & C_{11} & 0 & 0 & 0 \\
0    & 0    & 0    & C_{44} & 0 & 0 \\
0    & 0    & 0    & 0    & C_{44} & 0 \\
0    & 0    & 0    & 0    & 0      & C_{44}
\end{pmatrix}\ \, ,
\end{equation}
\begin{equation}
e_{nI}\leftrightarrow
\begin{pmatrix}
0 & 0 & 0 & e_{14} & 0 & 0 \\
0 &0 & 0 & 0 & e_{14} & 0 \\
0 & 0 &0 & 0 & 0 & e_{14}
\end{pmatrix} \, ,
\end{equation}
\begin{equation}
\epsilon_{mn} \leftrightarrow
\begin{pmatrix}
\epsilon_{11} & 0 & 0  \\
0 &\epsilon_{11} & 0  \\
0 & 0 &\epsilon_{11}
\end{pmatrix}\ \, ,
\end{equation}
where $C_{11}$, $C_{12}$, $C_{44}$, $e_{14}$, and $\epsilon_{11}$ are the only independent material constants. In the case of the crystal class $O_h$ corresponding to non-piezoelectric materials, $e_{14}$ must be taken to be zero.

If a rotated system of axes is to be used, then it is necessary to first transform accordingly the material tensors and only afterwards construct the associated Voigt matrices. This is the situation in Sec.~\ref{result}, where a nanowire with $\hat{X}_3$ axis along the crystallographic direction [111] is studied. For that purpose it is convenient to employ a new system of axes
($\hat{X}_1 \parallel [1 0 \bar{1}]$, $\hat{X}_2 \parallel [\bar{1}2\bar{1}]$, $\hat{X}_3 \parallel [111]$). The rotation matrix leading to
this new reference frame is:
\begin{equation}
R=
\begin{pmatrix}
\frac{1}{\sqrt{2}} & 0  &-\frac{1}{\sqrt{2}} \\
-\frac{1}{\sqrt{6}} & \frac{2}{\sqrt{6}}  & -\frac{1}{\sqrt{6}}\\
\frac{1}{\sqrt{3}}&\frac{1}{\sqrt{3}} &\frac{1}{\sqrt{3}} \\
\end{pmatrix}\ \, .
\end{equation}
Following the procedure outlined above, one obtains the following Voigt matrices
corresponding to the rotated system of axes:
\begin{equation}
\hat{C}_{IK} \leftrightarrow
\begin{pmatrix}
\hat{C}_{11} & \hat{C}_{12} & \hat{C}_{12} & \hat{C}_{14} & 0 & 0 \\
\hat{C}_{12} & \hat{C}_{11} & \hat{C}_{12} &-\hat{C}_{14} & 0 & 0 \\
\hat{C}_{12} & \hat{C}_{12} & \hat{C}_{33} & 0 & 0 & 0 \\
 \hat{C}_{14}   & -\hat{C}_{14}   & 0    & \hat{C}_{44} & 0 & 0 \\
0    & 0    & 0    & 0    & \hat{C}_{44} & \hat{C}_{14} \\
0    & 0    & 0    & 0    & \hat{C}_{14}     & \hat{C}_{66}
\end{pmatrix}\ \, ,
\end{equation}
  \begin{equation}
\hat{e}_{nI}\leftrightarrow
\begin{pmatrix}
0 & 0 & 0 & 0 &  \hat{e}_{15}  & -\hat{e}_{22}  \\
-\hat{e}_{22} &  \hat{e}_{22}  & 0 &  \hat{e}_{15}  & 0 & 0 \\
\hat{e}_{31}  & \hat{e}_{31} & \hat{e}_{33} & 0 & 0 & 0
\end{pmatrix}  \, ,
\end{equation}
\begin{equation}
\hat{\epsilon}_{mn} \leftrightarrow
\begin{pmatrix}
\epsilon_{11} & 0 & 0  \\
0 &\epsilon_{11} & 0  \\
0 & 0 &\epsilon_{11}
\end{pmatrix}\  \, ,
\end{equation}
where the elements of the matrices are:
\begin{alignat*}{2}
\hat{C}_{11}&= \frac{1}{2} (C _{11}+C _{12}+2C _{44})         \, ,   \\[1ex]
\hat{C}_{12}&= \frac{1}{6} (C _{11}+5C _{12}-2C _{44})        \, ,   \\[1ex]
\hat{C}_{13}&= \frac{1}{3} (C _{11}+2C _{12}-2C _{44})        \, ,   \\[1ex]
\hat{C}_{14}&= \frac{1}{3\sqrt{2}} (-C _{11}+C _{12}+2C _{44}) \, ,   \\[1ex]
\hat{C}_{33}&= \frac{1}{3} (C _{11}+2C _{12}+4C _{44})        \, ,   \\[1ex]
\hat{C}_{44}&= \frac{1}{3} (C _{11}-C _{12}+C _{44})          \, ,   \\[1ex]
\hat{C}_{66}&= \frac{1}{2} (\hat{C}_{11}-\hat{C}_{12})        \, ,
\end{alignat*}
\begin{alignat*}{2}
\hat{e}_{15}& = -\sqrt{\frac{1}{3}}\,e _{14}    \, , \\[1ex]
\hat{e}_{22}& = \hphantom{-} \sqrt{\frac{2}{3}}\,e _{14}      \, ,   \\[1ex]
\hat{e}_{31}&= -\sqrt{\frac{1}{3}}\,e _{14}  \, , \\[1ex]
\hat{e}_{33}&=  \hphantom{-}\sqrt{\frac{4}{3}}\,e _{14}    \, .
\end{alignat*}

\section{Strain and electric field in cylindrical coordinates}\label{APP:Cylindrical_components}

The relation between the cylindrical and Cartesian components of the strain tensor and
electric field vector are given as:
\begin{alignat*}{2}
\varepsilon_{rr} &= \varepsilon_{11}\;\cos^2\phi +\varepsilon_{22}\;\sin^2\phi+
\varepsilon_{12}\; \sin  2\phi  \, ,   \\[1ex]
\varepsilon_{\phi \phi} &= \varepsilon_{11}\; \sin^2\phi+\varepsilon_{22}\;\cos^2\phi-
\varepsilon_{12}\; \sin  2\phi  \, ,   \\[1ex]
\varepsilon_{zz} &= \varepsilon_{33} \, ,   \\[1ex]
\varepsilon_{r \phi} &= \frac{1}{2}(\varepsilon_{22}-\varepsilon_{11})\; \sin  2\phi+  \varepsilon_{12}\; \cos  2\phi\, ,   \\[1ex]
\varepsilon_{rz} &= \varepsilon_{23}\;\sin\phi+\varepsilon_{13} \; \cos\phi \, ,   \\[1ex]
\varepsilon_{\phi z} &=  \varepsilon_{23}\;\cos\phi-\varepsilon_{13}\;\sin\phi \, ,
\end{alignat*}

\begin{alignat*}{2}
E_{r} &= \hphantom{-}E_1\;\cos\phi +E_2\;\sin\phi  \, ,   \\[1ex]
 E_{\phi} &=  -E_1\;\sin\phi +E_2\;\cos\phi  \, ,   \\[1ex]
 E_{z} &= E_{3} \, .
\end{alignat*}

\end{appendices}

\newpage

\bibliography{bibliography}

\end{document}